\documentclass[aps,nofootinbib,superscriptaddress, showpacs,preprintnumbers,  nofootinbibt,twocolumn]{revtex4-1}
\usepackage{eurosym}
\usepackage{amsmath}
\usepackage{bm}
\usepackage{amsfonts}
\usepackage{amssymb}
\usepackage{graphicx}
\usepackage{mathtools}
\usepackage{color}

\def\be{\begin{equation}}
\def\ee{\end{equation}}
\def\bea{\begin{eqnarray}}
\def\eea{\end{eqnarray}}

\begin{document}

\title{Constraining Weyl type $f(Q,T)$ gravity with Big Bang Nucleosynthesis}
\author{Jian Ge}
\email{2672537261@qq.com}
\affiliation{ School of Physics,
Sun Yat-Sen University, Guangzhou 510275, People’s Republic of China,}
\author{Lei Ming}
\email{minglei@scnu.edu.cn}
\affiliation{ Key Laboratory of Atomic and Subatomic Structure and Quantum Control(Ministry of Education), Guangdong Basic Research Center of Excellence for Structure and Fundamental Interactons of Matter, School of Physics, South China Normal University, Guangzhou 510006, China}
\affiliation{ Guangdong Provincial Key Laboratory of Quantum Engineering and Quantum Materials, Guangdong-Hong Kong Joint Laboratory of Quantum Matter, South China Normal University, Guangzhou 510006, China}
\affiliation{ School of Physics,
Sun Yat-Sen University, Guangzhou 510275, People’s Republic of China,}
\author{Shi-Dong Liang}
\email{stslsd@mail.sysu.edu.cn}
\affiliation{ School of Physics,
Sun Yat-Sen University, Guangzhou 510275, People’s Republic of China,}
\author{Hong-Hao Zhang}
\email{zhh98@mail.sysu.edu.cn}
\affiliation{ School of Physics,
Sun Yat-Sen University, Guangzhou 510275, People’s Republic of China,}
\author{Tiberiu Harko}
\email{tiberiu.harko@aira.astro.ro}
\affiliation{Department of Physics, Babes-Bolyai University, Kogalniceanu Street,
	Cluj-Napoca, 400084, Romania,}
\affiliation{Astronomical Observatory, 19 Ciresilor Street,
	Cluj-Napoca 400487, Romania}

\begin{abstract}
The Weyl type $f (Q,T)$ modified gravity theory is an extension of the $f(Q)$ and $f(Q,T)$ type theories, where $T$ is the trace of the matter energy-momentum tensor, and the scalar non-metricity $Q$ is represented in its standard Weyl form, and it is fully determined by a vector field $\omega _\mu$.  The theory can give a good description of the observational data, and of the evolution of the late-time Universe, including a geometric explanation of the dark energy. In this work we investigate the Big Bang Nucleosynthesis (BBN) constraints on several Weyl type $f (Q,T)$ gravity models.  In particular, we consider the corrections that Weyl type $f(Q,T)$ terms induce  on the freeze-out temperature $\mathcal{T}_f$, as compared to the standard $\Lambda$CDM results. We analyze in detail three distinct cosmological models, corresponding to specific choices of the functional form of $f(Q,T)$. The first model has a simple linear additive structure in $Q$ and $T$, the second model is multiplicative in $Q$ and $T$, while the third is additive in $T$ and the exponential of $Q$.  For each $f(Q,T)$ we consider first the cosmological evolution in the radiation dominated era, and then we impose the observational bound on $\left|\delta \mathcal{T}_f/ \mathcal{T}_f\right|$ to obtain constraints on the model parameters  from the primordial abundances of the light elements such as helium-4, deuterium and lithium-7. The abundances of helium-4 and deuterium agree with theoretical predictions, however, the lithium problem, even slightly alleviated,  still persists for the considered Weyl type $f(Q,T)$ models. Generally, these models satisfy the BBN constraints, and thus they represent viable cosmologies describing the entire dynamical time scale of the evolution of the Universe.
\end{abstract}

%\pacs{03.75.Kk, 11.27.+d, 98.80.Cq, 04.20.-q, 04.25.D-, 95.35.+d}
\date{\today }
\maketitle
\tableofcontents

%\preprint{APS/123-QED}

\section{Introduction}

The theory of general relativity plays a vital role in modern physics and cosmology. Einstein \cite{Ein}  and Hilbert \cite{Hil} did extensively use the Riemannian geometry, by endowing the space-time with a metric, and an affine structure. The geometric and gravitational properties of the space-time can thus be described by a metric, and by the Riemann curvature tensor, and its contractions.

Shortly after the theory of general relativity was proposed, in 1918 Weyl introduced an extension of Riemannian geometry \cite{Weyl1, Weyl2}, with the goal to develop the first unified theory of gravity and electromagnetism. In Weyl's geometry, the presence of electromagnetic field can be attributed to the existence of the nonmetricity of the space-time. A few years later, Cartan \cite{Cart1,Cart2} introduced another generalization of the Riemann geometry, by considering another important geometric concept, the torsion tensor, describing the properties of a space-time with non-symmetric Christoffel symbols.

In its first 50 years of existence,  Weyl's geometry was largely ignored by physicists, mostly due to Einstein's strong criticism of the physical interpretation of the nonmetricity. But gradually the interest of physicists in this geometry begin to increase, and interesting physical and mathematical effects on both microscopic and macroscopic level have been intensively investigated \cite{Scholz}.

A number of recent theoretical investigations have arrived to the important result that general relativity can be described in three mathematically
equivalent geometric formalisms, by using the three basic geometric concepts: curvature, torsion, and non-metricity \cite{Jim}. Hence, one can consider, as a viable alternative to the curvature formulation, the teleparallel formulation, with the torsion vanishing for the first approach,  and the curvature vanishing for the second alternative, called the teleparallel equivalent of General Relativity (TEGR) \cite{TEGR, TEGR1}. The TEGR theory is also called the $f(\mathcal{T} )$ gravity
theory, and in it the torsion $\mathcal{T}$ exactly compensates curvature, with the space-time becoming flat.

The possibility of formulating general relativity by using only non-metricity was investigated in \cite{Nester}, leading to the development of the symmetric teleparallel formulation of gravity. This approach  was further generalized  into the $f(Q)$ gravity theory \cite{Jim1}, which is also called coincident general relativity, and in which $Q$ represents  quadratic nonmetricity scalar. The coupling of matter with nonmetricity was considered in \cite{fQ1}, and this initial study led to the extension of the symmetric teleparallel gravity to the $f(Q,T$) theory \cite{ref1}, in which the gravitational Lagrangian is given by an arbitrary  function $f$ of the non-metricity $Q$ and of the trace $T$ of the matter-energy-momentum tensor $T$.

The role of the modified theories of gravity has recently become central in the understanding of the gravitational dynamics. This situation was determined by the recent theoretical challenges faces by general relativity. These challenges come from a plethora of observational data, which cannot be explained using standard general relativity without adding at least one extra parameter to the theory.
The first major challenge general relativity is encountering is the explanation of the observations of the accelerated expansion of the Universe \cite{P1,P2,P3}, which is confirmed by data of the Planck satellite \cite{42}, and studies of Baryon Acoustic Oscillations \cite{43,44,45}. For a recent review of the
cosmic acceleration problem see \cite{46}.

The problem of the cosmic acceleration was addressed within the framework of the $f(Q,T)$ gravity \cite{ref1}, by obtaining first the cosmological evolution equations for a flat, homogeneous and isotropic geometry, which generalize the Friedmann equations of general relativity. Several cosmological models were investigated by adopting some simple functional forms for the function $f( Q, T)$: $f(Q,T)=\alpha  Q+\beta T$, $f(Q,T)=\alpha  Q^{n+1}+\beta  T$, and $f(Q,T)=-\alpha Q-\beta  T^2$, respectively. The main cosmological parameters (Hubble function, the deceleration parameter, and the matter-energy density) have been obtained as a function of the redshift by using analytical and numerical methods.

A particular class of the $f(Q,T)$ gravity theory was investigated in \cite{ref2}, in which the scalar non-metricity $Q_{\alpha \mu \nu}$ is represented  in a standard Weyl form, and it is fully determined by a vector field $\omega_\mu$. The field equations of the theory have been obtained by assuming the vanishing of the total scalar curvature, with this condition added into the gravitational action via a Lagrange multiplier. The gravitational field equations have been obtained from a variational principle, and they depend on the scalar nonmetricity, as well as on the Lagrange multiplier. The cosmological implications of the theory have been also investigated, and the cosmological evolution equations, which generalize the Friedmann equations of standard general relativity, have been obtained for a flat, homogeneous and isotropic geometry. Several cosmological models have been considered for some simple functional forms of $f(Q, T)$.  The predictions of the theory have been compared with the standard $\Lambda$CDM model, and the observational data. It turns out that the Weyl type $f(Q,T)$ gravity theory can give a good description of the observational data, and hence it can be considered as an attractive theoretical alternative to the standard $\Lambda$CDM model. Various physical and cosmological aspects of the $f(Q,T)$ type theory have been investigated in \cite{W1,W2,W3,W4,W5,W6,W7,W8,W9,W10,W11, W12, W13, W14,W15,W16,W17,W18, W19}.

 Big Bang Nucleosynthesis happened within the first three minutes of the Universe's existence. After the first $180\; \rm s$ from the Big Bang, the temperature of the Universe (defined as the radiation temperature) drops from beyond $10^{16}\; \rm GeV$ $= 1.2\times 10^{29}\; \rm K$ to $1\; \rm MeV$ $=1.2\times 10^{10}\; \rm K$, which represents the temperature at which the density of the nucleon component allowed for nuclear reactions to form stable nuclei \cite{Copi, Fields}. The presence of a small nucleon content allowed for the production of light nuclide, such as hydrogen, the most abundant element in the Universe, helium, the second most common element, and an isotope of lithium, respectively. The BBN processes took place during the rapid expansion of the Universe, and they lasted until the nuclear reactions ceased due to the low temperatures and matter densities, a result of the expansion of the Universe .
	
The standard model of cosmology heavily relies on the standard model of particle physics, thus incorporating many results of the elementary particle physics, like, for example,  the existence of three families of neutrinos. The formation of the light nuclei in the early Universe depend on several physical factors like temperature, nucleon density, expansion rate, neutrino content, and neutrino-antineutrino asymmetry, respectively. Therefore, BBN provides powerful methods to investigate and to test cosmological scenarios and physical conditions that might have influenced the formation of the primordial elements \cite{ Copi, Fields, Steigman-2004, Sepico, Steigman-2007, Steigman-2008, Li1, Steigman-2012, Li2, Cybrut, 2016On, F, De1,N1}.

BBN data were used in \cite{Barrow} to impose constraints on the exponent of the Barrow entropy, an extended entropy relation arising from the incorporation of quantum-gravitational effects on the black-hole structure.  Constraints on the parameters of the bimetric theory of gravity, a ghost-free and observationally viable extension of general relativity, exhibiting both a massless and a massive graviton, were obtained in \cite{N2}. Constraints obtained from BBN on a number of higher-order modified gravity theories, including the $f(G)$ Gauss-Bonnet gravity, the $f(P)$ cubic gravities, obtained through the use of the quadratic-curvature Gauss-Bonnet $G$ term, and the cubic-curvature combination, string-inspired quadratic Gauss-Bonnet gravity coupled to the dilaton field,  models with string-inspired quartic curvature corrections, and  running vacuum models were investigated in \cite{N3}.  It was found that all the above mentioned models can satisfy the BBN constraints, and hence they represent viable cosmological scenarios.  $f\left(\mathcal{T},\mathcal{T}_G\right)$ gravity, were $\mathcal{T}$ is the torsion scalar, and $\mathcal{T}_G$ is the teleparallel equivalent of the Gauss–Bonnet term, was analyzed with respect to the BBN constraints in \cite{N4}. The deviations of the freeze-out temperature determined  by the presence of the extra torsion terms were calculated, and a comparison to the standard $\lambda$CDM model was performed. It turns out that the BBN constraints are satisfied by a specific logarithmic model for large regions of the model parameters. The implications of Einstein-Aether and modified Horava-Lifshitz theories of gravity on the formation of light elements in the early Universe were considered in \cite{N5}. Some models in the framework of the considered theories can satisfy the Big Bang nucleosynthesis constraints. The BBN formalism and observations were used in \cite{Anag} to obtain constraints on several classes of $f(Q)$ gravity models. It was shown that models based on the $f(Q)$ gravity theory can easily satisfy the BBN constraints. This result indicates that $f(Q)$ modified gravity theories have an important advantage with respect to other modified gravity approaches. The BBN constraints on the cosmology of the bumblebee model, a vector-tensor theory of gravitation, were considered in \cite{N6}. In this model a single Lorentz-violating timelike vector field with a nonzero vacuum expectation value couples to the Ricci tensor and scalar.  By using the observational data coming from the BBN, and the matter-antimatter asymmetry in the Baryogenesis era,  some constraints on the vacuum expectation value  of the bumblebee timelike vector field can be imposed. The BBN was used in \cite{N7} to constrain the free parameters of the extended Starobinsky model. The primordial abundances of the light elements, such as He, D, He, or Li, have been compared with the most recent observational data. Most non-standard cosmic evolutions can not easily satisfy these BBN constraints, but a free parameter of the viable models has an upper limit imposed by the constraints.  BBN bounds on the model parameters have been obtained for the $f(R)$ gravity scalarons \cite{N8}, on the $f(T,\mathcal{T}$ gravity \cite{N9}, on the variation of the gravitational constant \cite{N10}, and on the quadratic energy–momentum squared gravity \cite{N11}, respectively.
	
It is the goal of this study to study the BIg Bang Nucleosynthesis (BBN) process within the framework of the Weyl type $f(Q,T)$ modified gravity theory. We use the Big Bang Nucleosynthesis (BBN) formalism and observed light element abundance to extract constraints on several $f(Q,T)$ models. The three models we are considering are given by $f(Q,T)=\alpha Q+\left(\beta /6\kappa ^2\right)T$, $f(Q,T)=\left(\alpha /6H_0^2\kappa ^2\right)QT$, and $f(Q,T)=\eta H_0^2\exp\left[\left(\mu /6H_0^2\right)\right]+\left(\nu /6\kappa ^2\right)T$, respectively, where $\alpha$, $\beta$, $H_0$, $\nu$, and $\eta$ are constants, while $\kappa ^2$ denotes the gravitational coupling constant. As a first step in our approach we obtain a general, although approximate,  expression for the ratio of the variation of the freeze-out temperature $\delta \mathcal{T}_f/\mathcal{T}_f$, by considering the corrections to the Hubble function due to the presence of the Weyl geometry effects. Moreover, a second approach is also used, in which we directly obtain the freeze-out temperature from the equality of the Hubble function and of the particle decay rate. For each model we begin our analysis of the BBN constraints by considering the full numerical evolution of the models during the radiation dominated period. We simultaneously analyze the behaviors of the Hubble functions of the models, of the particle decay rates, and of the Weyl vector. A comparison with the Hubble function behavior of standard general relativity is also performed. Then we explicitly obtain the BBN constraints on the model parameters from the observational values of the $^{4}$He mass fraction and $^{7}$Li abundance. The combined constraints coming from the observed numerical values of the $^{4}$He  $^{7}$Li are also considered. Based on the obtained results we can conclude that the Weyl type $f(Q,T)$ cosmological models can not only pass the cosmological tests in the early Universe, but they can also represent some viable and attractive alternatives for the physical description of the very early Universe.

The present paper is organized as follows. The theoretical formalism of the Weyl type $f(Q,T)$ gravity is briefly reviewed in Section~\ref{sect1}. The standard BBN theory, the theoretical freeze-out constraints, as well as the observational data and results are presented in Section~\ref{sect2}. The BBN constraints on three Weyl type $f(Q,T)$ are analyzed in detail in Section~\ref{sect3}. We discuss and conclude our results in Section~\ref{sect4}.

\section{Weyl type $f(Q,T)$ gravity}\label{sect1}

In the present Section we introduce the action of the Weyl type $f(Q,T)$ gravity, and we briefly review the basic geometric concepts necessary for the development of the basic theory. The field equations of the theory are obtained by varying the action with respect to the metric tensor. The generalized Friedmann equations, obtained for a flat FLRW geometry, are also presented.

\subsection{The gravitational field equations of the $f(Q,T)$ theory}

In the present Subsection we briefly present the action of the Weyl type $f(Q,T)$ gravity theory, as well the geometrical foundations of the theory.

\subsubsection{Action and geometry}

The action of the Weyl type $f(Q,T)$ gravity theory is given by \cite{ref2}
\bea
    \mathcal{S}&=&\int \left[\kappa^{2}f(Q,T)-\frac{1}{4}\Omega_{\mu\nu}\Omega ^{\mu\nu}-\frac{1}{2}m^{2}\omega^{2}+\mathcal{L}_{m}\right]\nonumber\\
   && \times \sqrt{-g}d^{4}x,
\eea
where $\kappa^{2}=1/16\pi G$, $Q$ is the non-metricity scalar, $\Omega_{\mu \nu}$ is the strength of the Weyl vector, while $\mathcal{L}_m$ is the action of the ordinary (baryonic) matter. This action is defined in a Weyl geometric framework, in which the metric condition is not satisfied, so that
\begin{equation}
Q_{\alpha\mu\nu}\equiv\bar{\nabla}_{\alpha}g_{\mu\nu}
    =2g_{\mu\nu}\omega_{\alpha},
\end{equation}
where $Q_{\alpha \mu \nu}$ is the general non-metricity tensor, and $\omega _\alpha$ is the Weyl vector, with square given by $\omega ^2=\omega _\alpha \omega ^\alpha$. The deformation tensor $L^{\lambda}_{\ \mu\nu}$ can be defined by using the non-metricity tensor as
\bea
L^{\lambda}_{\ \mu\nu}&\equiv & -\frac{1}{2}g_{\lambda\gamma}(Q_{\mu\gamma\nu}+Q_{\nu\gamma\mu}-Q_{\gamma\mu\nu})\nonumber\\
    &=&-\delta^{\lambda}_{\nu}\omega_{\mu}-\delta^{\lambda}_{\mu}\omega_{\nu}+g_{\mu\nu}\omega^{\lambda}.
\eea

The non-metricity scalar $Q$ is introduced according to the definition
\begin{equation}
\begin{split}
    Q&\equiv-g^{\mu\nu}(L^{\alpha}_{\ \beta\mu}L^{\beta}_{\ \nu\alpha}-L^{\alpha}_{\ \beta\alpha}L^{\beta}_{\ \mu\nu})
    =-6\omega^{2}.
\end{split}
\end{equation}

The strength of the Weyl vector is defined as
\begin{equation}
    \Omega_{\mu\nu}\equiv\nabla_{\nu}\omega_{\mu}-\nabla_{\mu}\omega_{\nu},
\end{equation}
where $\nabla_{\mu}$ denotes the covariant derivative with respect to Levi-Civita connection of the Riemannian geometry.

The Weyl scalar, the second contraction of the Weyl curvature tensor, is obtained as
\begin{equation}
    \bar{R}=R+6(\nabla_{\nu}\omega^{\nu}-\omega^{2}),
\end{equation}
where $R$ is the Ricci scalar of the Riemannian geometry.

In the following we impose the flatness condition, which requires $\bar{R}=0$. This condition is implemented in the action with the use of a Lagrangian multiplier $\lambda$. Hence, for the action of the Weyl type $f(Q,T)$ gravity we obtain the final expression
\begin{equation}
    \begin{split}
        \mathcal{S}&=\int d^{4}x\sqrt{-g}[\kappa^{2}f(Q,T)-\frac{1}{4}\Omega _{\mu\nu}\Omega ^{\mu\nu}-\frac{1}{2}m^{2}\omega^{2}\\
        &+\lambda(R+6\nabla_{\alpha}\omega^{\alpha}-6\omega^{2})+\mathcal{L}_{m}].
    \end{split}
\end{equation}

We also introduce the matter energy-momentum tensor, defined according to
\begin{equation}
    T_{\mu\nu}\equiv-\frac{2}{\sqrt{-g}}\frac{\delta(\sqrt{-g}\mathcal{L}_{m})}{\delta g^{\mu\nu}}=g_{\mu\nu}-2\frac{\delta\mathcal{L}_{m}}{\delta g^{\mu\nu}},
\end{equation}
The trace of the energy-momentum tensor is defined as $T=T_\mu^\mu$.

\subsubsection{Field equations}

By  considering the variation of the action with respect to the Weyl vector $\omega^{\mu}$, and by taking into account that $\delta  S/\delta \omega ^\mu=0$, after an integration by parts, we obtain the equation describing the variation of the Weyl vector as
\begin{equation}
 %   \nabla^{\nu}W_{\mu\nu}-(m^{2}+12\kappa^{2}f_{Q}+12\lambda)\omega_{\mu}=6\nabla_{\mu}\lambda,
 \nabla^{\nu}\Omega _{\mu\nu}-m_{eff}^2\omega_{\mu}=6\nabla_{\mu}\lambda,
\end{equation}
where we have denoted
\begin{equation}
    m_{eff}^{2}=m^{2}+12\kappa^{2}f_{Q}+12\lambda.
\end{equation}

Thus, in the present theory, the Weyl vector satisfies a Proca-type field equation. We also introduce the notations  $f_T=\partial f(Q,T)/\partial T$, and $f_q=\partial f(Q,T)/\partial Q$, respectively. The variation of the action with respect to the metric $g^{\mu\nu}$ yields the field equations \cite{ref2}
\bea
&& \frac{1}{2}(T_{\mu\nu}+S_{\mu\nu})-\kappa^{2}f_{T}(T_{\mu\nu}+\Theta_{\mu\nu})+\frac{\kappa^{2}}{2}g_{\mu\nu}f\nonumber\\
 &&=-6\kappa^{2}f_{Q}\omega_{\mu}\omega_{\nu}
       +\lambda(R_{\mu\nu}-6\omega_{\mu}\omega_{\nu}+3g_{\mu\nu}\nabla_{\rho}\omega^{\rho})\nonumber\\
      && +3g_{\mu\nu}\omega^{\rho}\nabla_{\rho}\lambda-6\omega_{(\mu}\nabla_{\nu)}\lambda
       +g_{\mu\nu}\Box\lambda-\nabla_{\mu}\nabla_{\nu}\lambda,
 \eea
where we have introduced the tensor $\Theta _{\mu \nu}$, defined  as
\begin{equation}
     \Theta_{\mu\nu}\equiv g^{\alpha\beta}\frac{\delta T_{\alpha\beta}}{\delta g^{\mu\nu}}=g_{\mu\nu}\mathcal{L}_{m}-2T_{\mu\nu}.
\end{equation}

The tensor $S_{\mu \nu}$ represents the energy-momentum tensor of the Proca field
\bea
            S_{\mu\nu}&\equiv&-\frac{1}{4}g_{\mu\nu}\Omega _{\rho\sigma}\Omega ^{\rho\sigma}+\Omega _{\mu\rho}\Omega _{\nu}^{\rho}-\frac{1}{2}m^{2}g_{\mu\nu}\omega^{2}
            +m^{2}\omega_{\mu}\omega_{\nu}.\nonumber\\
        \eea

 The gravitational equations of the Weyl type $f(Q,T)$ theory can be reformulated in a form similar to the standard Einstein field equations as
\bea\label{2}
       &&R_{\mu\nu}-\frac{1}{2}g_{\mu\nu}R=\frac{1}{2\lambda}(T_{\mu\nu}+S_{\mu\nu})-\frac{\kappa^{2}}{\lambda}f_{T}(T_{\mu\nu}+\Theta_{\mu\nu})\nonumber\\
       &&+\frac{\kappa^{2}}{2\lambda}g_{\mu\nu}f+\frac{6\kappa^{2}}{\lambda}f_{Q}\omega_{\mu}\omega_{\nu}-\frac{3}{\lambda}g_{\mu\nu}\omega^{\rho}\nabla_{\rho}\lambda+\frac{6}{\lambda}\omega_{(\mu}\nabla_{\nu)}\lambda\nonumber\\
       &&-\frac{1}{\lambda}(g_{\mu\nu}\Box\lambda-\nabla_{\mu}\nabla_{\nu}\lambda)+6\omega_{\mu}\omega_{\nu}-3g_{\mu\nu}\omega^{2}.
\eea

From the trace of the field equations we find
\bea
            \frac{1}{2}(T+S)-\kappa^{2}f_{T}(T+\Theta)&=&-2\kappa^{2}f-6\kappa^{2}f_{Q}\omega^{2}\nonumber\\
            &-&\frac{m^{2}}{2}\nabla^{\mu}\omega_{\mu}
            -6\kappa^{2}\nabla^{\mu}(f_{Q}\omega_{\mu}), \nonumber\\
        \eea
where $S=S_\mu^\mu=-m^{2}\omega^{2}$

The covariant derivative of both sides of Eq.~(\ref{2}) gives
\bea\label{16}
    \left(\frac{1}{2}+\kappa^{2}f_{T}\right)\nabla^{\mu}T_{\mu\nu}&=&\kappa^{2}g_{\mu\nu}\nabla^{\mu}(f_{T}\mathcal{L}_{m})-\kappa^{2}T_{\mu\nu}\nabla^{\mu}f_{T}\nonumber\\
    &&-\frac{\kappa^{2}}{2}f_{T}\nabla_{\nu}T.
\eea

From Eq.~(\ref{16}) it follows that due to the geometry-matter interaction, in the Weyl type $f(Q,T)$ theory an extra-force, acting on the massive test particle, is generated. The extra force is given by \cite{ref2}
\bea
        \mathcal{F}^{\rho}&=&-\frac{h^{\nu\rho}\nabla_{\nu}p}{\rho+p}+\frac{h^{\nu\rho}\nabla^{\mu}T_{\mu\nu}}{\rho+p}\nonumber\\\
        &=&\frac{\kappa^{2}h^{\nu\rho}}{(\rho+p)(1+2\kappa^{2}f_{T})}\Big[-\frac{1}{\kappa^{2}}\nabla_{\nu}p+2\nabla_{\nu}\left(f_{T}\mathcal{L}_{m}\right)\nonumber\\
        &&-f_{T}\nabla_{\nu}T
        -2T_{\mu\nu}\nabla^{\mu}f_{T}\Big],
\eea
where by $h^{\mu \nu}=g^{\mu \nu}+u^\mu u^\nu$ we have denoted the projection operator.

\subsection{The generalized Friedmann equations}

To investigate Big Bang Nucleosynthesis in the Weyl type $f(Q,T)$ gravity theory we assume first that the early Universe can be described by the flat, homogeneous and isotropic Friedmann-Lemaitre-Robertson-Walker (FLRW) metric, given by
\begin{equation}
    ds^{2}=-dt^{2}+a^{2}(t)\delta_{ij}dx^{i}dx^{j},
\end{equation}
where $a(t)$ is the scale factor. We also define the Hubble function as $H=\dot{a}/a$. In the following by a dot we denote the derivative with respect to the cosmological time $t$. For the Weyl vector we adopt the particular form
\begin{equation}
    \omega_{\mu}=(\psi(t),0,0,0)
\end{equation}
which follows from the conditions of the homogeneity and isotropy of the Universe.  With this choice we easily obtain the relations $\omega^{2}=-\psi^{2}$, and $Q=6\psi ^2$, respectively.

Hence, the energy balance equation and the generalized Friedmann equations of the Weyl type $f(Q,T)$ gravity can be written down as
\begin{equation}\label{eba}
        \dot{\rho}+3H(\rho+p)
        =\frac{-\kappa^{2}}{1+2\kappa^{2}f_{T}}[2(\rho+p)\dot{f}_{T}+f_{T}(\dot{\rho}-\dot{p})],
 \end{equation}
\bea\label{Fr1}
        3H^{2}&=&\frac{1}{2\lambda}[(1+2\kappa^{2}f_{T})\rho+2\kappa^{2}f_{T}p-\kappa^{2}f-\frac{m^{2}\psi^{2}}{2}\nonumber\\
        &&-6\lambda\psi^{2}
        +m_{eff}^{2}H\psi]=\frac{1}{2\lambda}(\rho+\rho_{eff}),
    \eea
and
\bea
2\dot{H}&=&\frac{-1}{2\lambda}[(1+2\kappa^{2}f_{T})(\rho+p)-\frac{m_{eff}^{2}}{3}(\dot{\psi}+\psi^{2}-H\psi)\nonumber\\
&&-4\kappa^{2}\dot{f}_{Q}\psi]
        =\frac{-1}{2\lambda}(\rho+\rho_{eff}+p+p_{eff}),
\eea
respectively, where we have introduced the notations
\begin{equation}
    \rho_{eff}\equiv m_{eff}^{2}H\psi+2\kappa^{2}f_{T}(\rho+p)-\kappa^{2}f-\frac{m^{2}\psi^{2}}{2}-6\lambda\psi^{2},
\end{equation}
and
\bea
         p_{eff}&\equiv&-\frac{m_{eff}^{2}}{3}(\dot{\psi}+\psi^{2}-H\psi)-4\kappa^{2}\dot{f}_{Q}\psi \nonumber\\
         &&-\left(m_{eff}^{2}H\psi
         -\kappa^{2}f-\frac{m^{2}\psi^{2}}{2}-6\lambda\psi^{2}\right)\nonumber\\
         &=&-\frac{m_{eff}^{2}}{3}(\dot{\psi}+\psi^{2}+2H\psi)-4\kappa^{2}\dot{f}_{Q}\psi+\kappa^{2}f\nonumber\\
         &&+\frac{m^{2}\psi^{2}}{2}+6\lambda\psi^{2},
\eea
respectively.

\section{Standard $\Lambda$CDM Big Bang Nucleosynthesis}\label{sect2}

 We begin our analysis of the Big Bang Nucleosynthesis constraints on the Weyl type $f(Q,T)$ gravity theory by briefly presenting the standard model of Big Bang Nucleosynthesis, developed in the framework of standard general relativity, and of the $\Lambda$CDM cosmological model. The standard Big Bang Nucleosynthesis model makes some definite predictions on the abundances of primordial elements, which we will review below.

\subsection{The radiation dominated Universe}

In the standard $\Lambda$CDM model, the two Friedmann equations describing the cosmological evolution are given by \cite{ref4}
\begin{equation}\label{hgr}
        3H_{GR}^{2}=\frac{\rho}{2\kappa^{2}},
\end{equation}
and
\begin{equation}\label{dhgr}
        2\dot{H}_{GR}=-\frac{1}{2\kappa^{2}}(\rho+p),
\end{equation}
respectively, where by $H_{GR}$ we have denoted the standard general relativistic expression of the Hubble function.  The matter content of the very early Universe is assumed to consist mainly of radiation, satisfying the equation of state $p_r=\rho _r/3$, and with energy density given by
\begin{equation}
    \rho\simeq\rho_{r}=\frac{\pi^{2}g_{\ast}}{30}\mathcal{T}^{4}
\end{equation}
where $\mathcal{T}$ is the temperature of the photon gas, and the effective number of degrees of freedom $g_{\ast}=10.6$. The evolution of the Hubble function is thus fully determined by the temperature of the photon gas
\begin{equation}\label{3}
    H_{GR}=\sqrt{\frac{\pi^{2}g_{\ast}}{180\kappa^{2}}}\mathcal{T}^{2}.
\end{equation}

In the radiation-dominated epoch, from the energy conservation equation we can easily find the relation
\begin{equation}
    \rho=\frac{\rho_{0}a_{0}^{4}}{a^{4}},
\end{equation}
where $\rho_0$ and $a_0$ are constants of integration. Combining this relation with Eq.~(\ref{hgr}), after integrating on both sides, we obtain
\begin{equation}
    a=\sqrt{\frac{\rho_{0}a_{0}^{4}}{6\kappa^{2}}t},
\end{equation}
and
\begin{equation}\label{time}
     \frac{1}{t}=\sqrt{\frac{2\rho_{r}}{3\kappa^{2}}}=\sqrt{\frac{\pi^{2}g_{\ast}}{45\kappa^{2}}}\mathcal{T}^{2},
\end{equation}
respectively.

\paragraph{The photon mass correction.} In the radiation-dominated epoch, we can consider the photon mass correction in the expression of the pressure of the radiation fluid. In the de Broglie-Proca cosmology \cite{2016de}, the pressure of the photons is given by
\begin{equation}
p_{r}=\frac{\rho_{r}}{3}\left(1-\frac{\sqrt{15g_\ast}m_{\gamma}^{2}}{6\pi\sqrt{2\rho_{r}}}\right)
        =\frac{\rho_{r}}{3}-\frac{\sqrt{15g_\ast}m_{\gamma}^{2}}{18\sqrt{2}\pi}\sqrt{\rho_{r}},
\end{equation}
with
$m_{\gamma}\leq10^{-18}\;{\rm eV}$. Hence,  in the radiation-dominated epoch, the radiation pressure can be expressed as
\begin{equation}
            p_r=\frac{\rho_{r}}{3}-\frac{g_\ast}{2}\dfrac{m_{\gamma}^{2}\mathcal{T}^{2}}{18}.
\end{equation}

For the trace of the energy-momentum tensor of the Proca type radiation field we obtain $T_r=-\rho_r+3p_r=-\left(g_\ast/2\right)m_\gamma ^2 \mathcal{T}^2/6$.
As the photon mass is small as compared with the Plank mass, in some physical situations we can directly ignore the last term in the expression of the pressure, and approximate it as $p_r=\rho_{r}/3$.

\subsection{The freeze-out constraint}

An important observational quantity in the study of the Big Bang Nucleosynthesis is the freeze-out temperature $ \mathcal{T}_f $. A fundamental physical process in the early Universe is the conversion of protons and neutrons taking place through the weak interactions. If the temperature of the Universe is greater than $ 1\;{\rm MeV} $,  the neutron-proton conversion reactions are in equilibrium. Once the temperature of the Universe becomes lower than  $ 1\;{\rm MeV} $, the neutron-to-proton ratio freezes out at a value of around $ 1/6 $, and then it slowly decrease due to free neutron decay \cite{ref4}. Consequently, the neutron abundance can be estimated once the neutron to proton $ \Gamma_{pn}(\mathcal{T}) $ and the proton to neutron $ \Gamma_{np}(\mathcal{T}) $ conversion rates are known.

Neutrons decay into protons  through three distinct channels \cite{Cybrut, ref4}
\begin{equation*}
 n\longrightarrow p+e^-+\Bar{\nu}_e ,
 \end{equation*}
 \begin{equation*}
  n+\nu_e\longrightarrow p+e^-
  \end{equation*}
   and
   \begin{equation*}
   n+e^+\longrightarrow p+\Bar{\nu}_e ,
   \end{equation*}
 respectively. Hence
\bea\label{25}
    \Gamma_{pn}(T)&=&\Gamma_{(n\longrightarrow p+e^-+\Bar{\nu}_e)}+\Gamma_{(n+\nu_e\longrightarrow p+e^-)}\nonumber\\
    &&+\Gamma_{(n+e^+\longrightarrow p+\Bar{\nu}_e)}.
\eea

Once the inverse decay rate $ \Gamma_{np}(\mathcal{T}) $ is known, the total conversion rate $ \Gamma_{tot}(\mathcal{T})=\Gamma_{pn}(\mathcal{T})+\Gamma_{np}(\mathcal{T}) $ can be obtained as \cite{Kolb, Bern}
\begin{equation}\label{26}
    \Gamma_{tot}(\mathcal{T})\equiv \Gamma (\mathcal{T})=4A\mathcal{T}^3(4!\mathcal{T}^2+2\times 3!\mathcal{Q}\mathcal{T}+2!\mathcal{Q}^2)
\end{equation}
here $\mathcal{Q}=m_n-m_p=1.29\times10^{-3}\; {\rm GeV}$ and $ A=1.02\times 10^{-11}\; {\rm GeV}^{-4} $.

When the temperature is large, the total conversion rate is much larger than the Hubble rate, and thus the protons and the neutrons are in the equilibrium. This equilibrium situation is maintained until a freeze-out temperature is reached, when the two particles decouple from each other.

The freeze-out temperature is defined in terms of the Hubble function according to the relation \cite{Kolb, Bern}
\be
 H(\mathcal{T}_f)=\Gamma_{tot}(\mathcal{T}_f)\approx c_q\mathcal{T}_f^5,
 \ee
 where $ c_q=4A4!\approx9.8\times 10^{-10}\;{\rm GeV}^{-4} $. Using Eq.~(\ref{3}) at $ \mathcal{T}=\mathcal{T}_f $, we find
\begin{equation}\label{27}
    \mathcal{T}_f=\bigg(\frac{\pi^2 g_*}{180\kappa ^2c_q^2}\bigg)^{1/6}\approx 0.6786\;{\rm MeV}.
\end{equation}

At the freeze-out time, the neutron-proton ratio is given as
\begin{equation}
    \left(\frac{n}{p}\right)_{f}\simeq e^{-\frac{\mathcal{Q}}{\mathcal{T}}}
\end{equation}
We define the neutron mass fraction as
\begin{equation}
    X\equiv\frac{n}{n+p}.
\end{equation}

The evolution equation of $X$ is given by \cite{ref4}
\begin{equation}\label{eqev}
    \frac{dX}{dx}=\frac{x\Gamma_{np}}{\sqrt{\frac{4\pi^{3}G\mathcal{Q}^{4}g_{\ast}}{45}}}[e^{-x}-X(1+e^{-x})].
\end{equation}
By substituting  $X$ with e$^{-\mu(x)}$X', where
\bea
  \mu(x)&\equiv&\int_{x_{i}}^{x}\frac{x'\Gamma_{np}(x')}{\sqrt{\frac{4\pi^{3}G\mathcal{Q}^{4}g_{\ast}}{45}}}(1+e^{-x'})dx' =-\frac{255}{\tau_{n}\sqrt{\frac{4\pi^{3}G\mathcal{Q}^{4}g_{\ast}}{45}}} \nonumber\\
    &\times& \left. \left[[\left(\frac{4}{x^{3}}+\frac{3}{x^{2}}+\frac{1}{x}\right)+\left(\frac{4}{x^{3}}+\frac{1}{x^{2}}\right)e^{-x}\right]\right|_{x_{i}}^{x},
\eea
with $x_{i}$ denoting the value of $x$ corresponding to a very  early time, $X\left(x_{i}\right)=X'\left(x_{i}\right)=0.5$, we can obtain
the solution of Eq.~(\ref{eqev}) as
%\begin{equation}
%    X'=0.5+\int_{x_{i}}^{x}\frac{x'\lambda_{np}(x')e^{-x'}}{\sqrt{\frac{4\pi^{3}G\mathcal{Q}^{4}g_{\ast}}{45}}}e^{\mu(x')}dx',
%\end{equation}
\begin{equation}
     X(x)=0.5e^{-\mu(x)}+\int_{x_{i}}^{x}\frac{x'\Gamma_{np}(x')e^{-x'}}{\sqrt{\frac{4\pi^{3}G\mathcal{Q}^{4}g_{\ast}}{45}}}e^{\mu(x')-\mu(x)}dx'.
\end{equation}
Then we obtain $X(\infty)=0.148$. The value of $X$ in the standard BBN is also represented as
\begin{equation}
    X_{f}=\frac{e^{-\frac{\mathcal{Q}}{\mathcal{T}_{f}}}}{1+e^{-\frac{\mathcal{Q}}{\mathcal{T}_{f}}}}=X(\infty),
\end{equation}
from which we obtain $(n/p)_{f}=0.1737$, and $\mathcal{T}_{f}\approx 0.7387$ MeV,

Between the freeze-out and the nucleosynthesis phases, a fraction of neutrons decay into protons
\begin{equation}
    X_{n}(t_{N})=X_{f}e^{-\frac{t_{N}-t_{f}}{\tau_{n}}}=\chi\frac{e^{-\frac{\mathcal{Q}}{\mathcal{T}_{f}}}}{1+e^{-\frac{\mathcal{Q}}{\mathcal{T}_{f}}}},
\end{equation}
where
$\chi\equiv e^{-\frac{t_{N}-t_{f}}{\tau_{n}}}$, thus
reducing the neutron mass fraction further. Supposing all neutrons have transformed  into $^{4}He$ after the BBN, the $^{4}He$ mass fraction is given as \cite{ref4}
\begin{equation}\label{He,theo}
	    Y_{P}=2X_{n}(t_{N})=\chi\frac{2e^{-\frac{\mathcal{Q}}{\mathcal{T}_{f}}}}{1+e^{-\frac{\mathcal{Q}}{\mathcal{T}_{f}}}}.
\end{equation}

\paragraph{The freeze-out temperature constraint.} Let's assume now that the freeze-out temperature can vary due to the presence of modified gravity, or other physical effects. Since $ H(\mathcal{T}_f)\approx c_q\mathcal{T}_f^5 $, we obtain immediately the relation $ \delta H\left(\mathcal{T}_f\right)=5c_q\mathcal{T}_f^4\delta \mathcal{T}_f $, which gives the relation
\begin{equation}
   \frac{\delta H\left(\mathcal{T}_f\right)}{H\left(\mathcal{T}_f\right)} =5\frac{\delta \mathcal{T}_f}{\mathcal{T}_f}.
\end{equation}
%The observational constraint for the variation of the freeze-out temperature is (see \cite{Anag} and references therein)
%\begin{equation}
%    \bigg|\frac{\delta \mathcal{T}_f}{\mathcal{T}_f}\bigg| < 4.7\times 10^{-4}.
%\end{equation}

The magnitude of the freeze-out constraint can be obtained as follows. From Eq.~(\ref{time}), we obtain first
\begin{equation}
    \frac{dt_{f}}{d\mathcal{T}_{f}}=-\frac{2t_{f}}{\mathcal{T}_{f}}.
\end{equation}
We take now the derivative of Eq.~(\ref{He,theo}) to find
\bea
 \frac{dY_{P}}{d\mathcal{T}_{f}}&=&-\frac{2t_{f}}{\tau_{n}\mathcal{T}_{f}}Y_{P}+\frac{\mathcal{Q}}{\mathcal{T}_{f}^{2}(1+x)}Y_{P}\nonumber\\
        &=&-\frac{2t_{f}}{\tau_{n}\mathcal{T}_{f}}Y_{P}+\left(1-\frac{Y_{P}}{2\chi}\right)\ln(\frac{2\chi}{Y_{P}}-1)\frac{Y_{P}}{\mathcal{T}_{f}}.
 \eea
Hence the deviation of the $^{4}He$ abundance can be represented as the deviation of the freeze-out temperature
\begin{equation}
    \delta Y_{P}=Y_{P}\left[\left(1-\frac{Y_{P}}{2\chi}\right)\ln\left(\frac{2\chi}{Y_{P}}-1\right)-\frac{2t_{f}}{\tau_{n}}\right]\frac{\delta \mathcal{T}_{f}}{\mathcal{T}_{f}}.
\end{equation}
From the observational data (see \cite{Anag} and references therein), we obtain for the deviation of the freeze-out temperature the upper limit
\begin{equation}\label{5}
    \left| \frac{\delta \mathcal{T}_{f}}{\mathcal{T}_{f}}\right| <4.7\times10^{-4}.
\end{equation}

As one can see, the deviation of the $^{4}$He abundance has a strong dependence on the freeze-out temperature, whose form does not change significantly in the non standard models. Hence,  we can reformulate the observations of the $^{4}He$ abundance into the constraints on the freeze-out temperature $\mathcal{T}_{f}$. Thus,  we do not need to use explicitly the value $Y_{P}$. We just need to make the freeze-out temperature not to be too large or too small to obtain the correct expression of the $^{4}$He abundance, whatever is the specific form of the freeze-out temperature.

With the degrees of freedom in the freeze-out temperature taken as  $g_{\ast}=10.6$ \cite{2016On}, we obtain the freeze-out time as $t_{f}=1.36$ s.
From the observational data, we have for the $He^{4}$mass fraction the value
    $Y_{P}=0.245$, and $\left| \delta Y_{P}\right|<2\times10^{-3}$. Combining this relation with the freeze-out time, we have
$t_{N}=169\;{\rm s}=2.57\times10^{26}\;{\rm GeV}^{-1}$.

By assuming for the number of the effective degree of freedom the value $g_{\ast}=3.37$, we obtain the nucleosynthesis temperature of the deuterium as
$\mathcal{T}_{N}=8.83\times10^{-2}$ MeV.

In the following we introduce the notation
\begin{equation}
	Z\equiv\frac{H}{H_{GR}}.
\end{equation}

\subsection{$^{2}H$, $^{4}He$ and $^7 Li$ abundances}

\paragraph{$^{2}H$.} Deuterium $^{2}H$ is produced in the early Universe through the reaction $n+p \rightarrow ^{2}H +\gamma$ \cite{ref4}. The best fit of the deuterium abundance is obtained as  \cite{Steigman-2012}
\begin{equation}\label{aDe}
Y_{D_ p} = 2.6\times (1 \pm 0.06) \times \left[ \frac{6}{\eta_{ 10} - 6 (Z-1)}\right]^{1.6}
\end{equation}
For $Z=1$ and $\eta_{ 10} = 6$, we obtain $y_{D_ p} = 2.6 \pm 0.16$. The observational constraint on the deuterium abundance is $y_{D p} = 2.55 \pm 0.03$ \cite{De1}.
Hence $Z$ is constrained in the range $Z=1.062 \pm 0.444$ \cite{N9}.

\paragraph{$^{4}He$.} The production of $^{4}He$ begins with the production of $^{2}H$ from the reaction of a  neutron $n$ and of a proton $p$. Then deuterium is converted into $^{3}He$ and tritium $T$ via the reactions $n + p \rightarrow ^{2}H + \gamma$,  $ ^{2}H+^{2}H \rightarrow ^{3}He + n$, and $ ^{2}H + ^{2}H \rightarrow ^{3}H + p$, respectively.  Finally, $^{4}He$ is produced from the reactions of $^{3}H$ with $^{2}H$ and $^{3}He$, given by
$^{2}H + ^{3}H \rightarrow ^{4}He + n$ and $ ^{2}H + ^{3}He\rightarrow ^{4}He + p$ \cite{Kolb, ref4}.

The cosmological $^{4}He$ abundance is obtained from the numerical best fit of the observational data as  \cite{Steigman-2004,Steigman-2007}
\begin{equation}
Y_{^{4}He} = 0.2485 \pm 0.0006 + 0.0016 \left[\left( \eta_{ 10} - 6\right) +100\left( Z-1\right)  \right].
\end{equation}
For $Z=1$, corresponding to standard Big Bang Nucleosynthesis, we obtain  $Y_{^{4}He} = 0.2485 \pm 0.0006$.
The cosmological observations give for $^{4}He$ abundance the value $0.2449 \pm 0.0040$ \cite{De1}. This constrains $Z$ in the range  $1.0475 \pm 0.105$ \cite{N9}.

\paragraph{$^{7}Li$.} The lithium abundance is also an important cosmological indicator, and the study of the abundance of these nuclei could provide a strong constraint on the theoretical model. The lithium abundance is given by the following numerical best fit \cite{Li1, Li2, Steigman-2012}
\begin{equation}\label{Li,num}
	\frac{^7 Li}{H}=4.82\times 10^{-10} \times (1\pm0.1)\times \left[\frac{10^{10}\eta-3(Z-1)}{6}\right]^2.
\end{equation}

In the standard BBN case, $Z=1$, and we obtain
\begin{equation}
	\frac{^7 Li}{H}=4.82\times 10^{-10}\times (1\pm 0.1)\times \left(\frac{10^{10}\eta}{6}\right)^2.
\end{equation}
After using the observed value of baryon-photon ratio
$\eta=6\times10^{-10}$,
we can obtain the ratio
\begin{equation}
    \frac{^7Li}{H}=4.82\times 10^{-10}\times  \left(1\pm 0.1\right),
\end{equation}
which is much larger than the observed value
\begin{equation}\label{Li,obs}
   \left. \frac{^7 Li}{H}\right|_{obs}=(1.6\pm0.3)\times 10^{-10}.
\end{equation}

The lithium abundance is important since the $\eta_{ 10}$ parameter, which fits well the abundances of hydrogen and helium, cannot describe the observations of $^{7}Li$. The ratio of the value of $^{7}Li$ abundance predicted by the standard Big Bang Nucleosynthesis model and the observations is of the order of $2.4-4.3$ \cite{Li1,Li2}. Hence, the standard BBN cannot explain the observed abundance of $^{7}Li$. This problem is known as the Lithium problem \cite{Li1}.
The observational value of the lithium abundance is $Y_{p}^{(Li)} = 1.6 \pm 0.3$ \cite{De1}. Thus, the constraint on $Z$ obtained from the observed $^{7}Li$ abundance is $Z = 1.960025 \pm 0.076675$. It is important to note that this constraint does not overlap with the constraints on Deuterium-2 and Helium-4 \cite{N9}.

\section{Nucleosynthesis constraints on Weyl type $f(Q,T)$ gravity models}\label{sect3}

In the present Section we  consider the BBN constraints on several Weyl type $f(Q,T)$ gravity cosmological models. We begin our analysis by imposing the flatness condition, which allows to express the time component Weyl vector as a function  of $H$. Then, we will obtain the general expression of the freeze-out temperature in Weyl type $f(Q,T)$ gravity, and we will apply these general equations to constrain some particular cosmological models.

\subsection{Freeze-out temperature in Weyl type $f(Q,T)$ gravity}

The flatness condition (zero total Weyl curvature $\tilde{R}$), introduced in the action via a Lagrange multiplier, can be formulated in the cosmological case as
\begin{equation}\label{psi}
\dot{H}+2H^{2}-\dot{\psi}-3H\psi+\psi^{2}=0,
\ee
or, equivalently
\be
\left(\dot{\psi}-\dot{H}\right)-(\psi-2H)(\psi-H)=0.
\end{equation}
The flatness condition is identically satisfied by the choice
\begin{equation}
    \psi=H,
\end{equation}
and in the following we will adopt, in the approximate analysis of the cosmological implications of our models,  this expression for $\psi$. We will also adopt for the Lagrange multiplier $\lambda$ the standard general relativistic approximation $\lambda \approx \kappa ^2 ={\rm constant}$. Moreover, in these approximations we obtain for the non-metricity scalar $Q$ the expression
\be
Q=6H^2.
\ee

From Eq.~(\ref{Fr1}) we obtain for the Hubble function the general expression
\be
H=\sqrt{\frac{1}{6\lambda }}\sqrt{\rho _{r}}\sqrt{1+\frac{\rho _{eff}}{\rho
_{r}}}.
\ee

By performing a first order series expansion of the square root we obtain
\begin{equation}\label{47}
\delta H=H-H_{GR}=\frac{1}{2}\frac{\rho _{eff}}{\rho _{r}}H_{GR}.
\end{equation}

Hence, for the variation of the freeze-out temperature we obtain the general
expression
\begin{equation}\label{48}
\left|\frac{\delta \mathcal{T}_{f}}{\mathcal{T}_{f}}\right|=\frac{1}{10c_{q}}\frac{1}{\mathcal{T}_{f}^{5}}\left.\left[ \frac{%
\rho _{eff}\left(\mathcal{T}\right)}{\rho _{r}\left(\mathcal{T}\right)}H_{GR}\left(\mathcal{T}\right)\right] \right |_{\mathcal{T}=\mathcal{T}_{f}}.
\end{equation}

Under the considered approximations we obtain successively
\be
m_{eff}^2=m^2+12\kappa ^2\left(1+f_Q\right),
\ee
and
\bea\label{50}
\rho_{eff}&=&-\kappa ^2f+2\kappa ^2f_T\left(\rho_r+p_r\right)\nonumber\\
&&+\frac{1}{2}H^2\left[m^2+12\kappa ^2\left(1+2f_Q\right)\right],
\eea
respectively. In Eq.~(\ref{50}) we can approximate $H$ by its general relativistic expression $H_{GR}$, $H\approx H_{GR}$, and $\rho_r$ and $p_r$ by their standard expressions, by ignoring the corrections due to the photon mass. Moreover, we have $H_{GR}^2/\rho_r=1/6\kappa ^2$. Hence we obtain
\be
\frac{\rho_{eff}}{\rho_r}=1-\kappa ^2 \frac{f}{\rho_r}+2f_Q+\frac{8}{3}\kappa ^2f_T+\frac{m^2}{12\kappa ^2}.
\ee

Thus, for the freeze-out constraint in the Weyl type $f(Q,T)$ gravity we obtain the general expression
\bea\label{53}
\left |\frac{\delta \mathcal{T}_f}{\mathcal{T}_f}\right|&=&\frac{1}{10c_q}\sqrt{\frac{\pi^2g_{\ast}}{180\kappa ^2}}\frac{1}{\mathcal{T}_f^3}\Bigg(1-\frac{30\kappa ^2}{\pi ^2g_{\ast}}\frac{f}{\mathcal{T}_f^4}+2f_Q\nonumber\\
&&+\frac{8}{3}\kappa ^2f_T
+\frac{m^2}{12\kappa ^2}\Bigg).
\eea

As for the quantity $Z=H/H_{GR}$, from Eq.~(\ref{47}) we obtain the expressions
\bea
\hspace{-0.5cm}Z&=&1+\frac{1}{2}\frac{\rho_{eff}}{\rho_r}\nonumber\\
\hspace{-0.5cm}&=&1+\frac{1}{2}\left(1-\kappa ^2 \frac{f}{\rho_r}+2f_Q+\frac{8}{3}\kappa ^2f_T+\frac{m^2}{12\kappa ^2}\right).
\eea

\subsection{Model I: $f(Q,T)=\alpha Q+\frac{\beta}{6\kappa^{2}}T$}

As a first example of a specific cosmological model in $f(Q,T)$ gravity, we consider the case when the function $f$ is given by
\begin{equation}
    f(Q,T)=\alpha Q+\frac{\beta}{6\kappa^{2}}T,
\end{equation}
where $\alpha $ and $\beta $ are constants. In this case $f_Q=\alpha$, and $f_T=\beta /6\kappa ^2$.

\begin{equation}
    H^{2}=\frac{1}{108\lambda}[(18+8\beta)\rho+18m_{eff}^{2}H\psi-9m_{eff}^{2}\psi^{2}],
\end{equation}
and
\begin{equation}
    \dot{H}=\frac{-1}{12\lambda}\Bigg[\frac{4}{3}(3+\beta)\rho+m_{eff}^{2}(-\dot{\psi}-\psi^{2}+H\psi)\Bigg],
\end{equation}
respectively. In the following we assume that $\lambda$ is a constant, $\lambda\equiv\kappa^{2}$. Moreover, we approximate $Q=6H^2$ as $Q\approx 6H_{GR}^2=\rho_r/\kappa ^2$. For $T$ we adopt the expression $T=T_r=-\left(g_\ast/2\right)m_\gamma ^2 \mathcal{T}^2/6$.

\begin{figure*}[ht]
\centering
\includegraphics[scale=0.3]{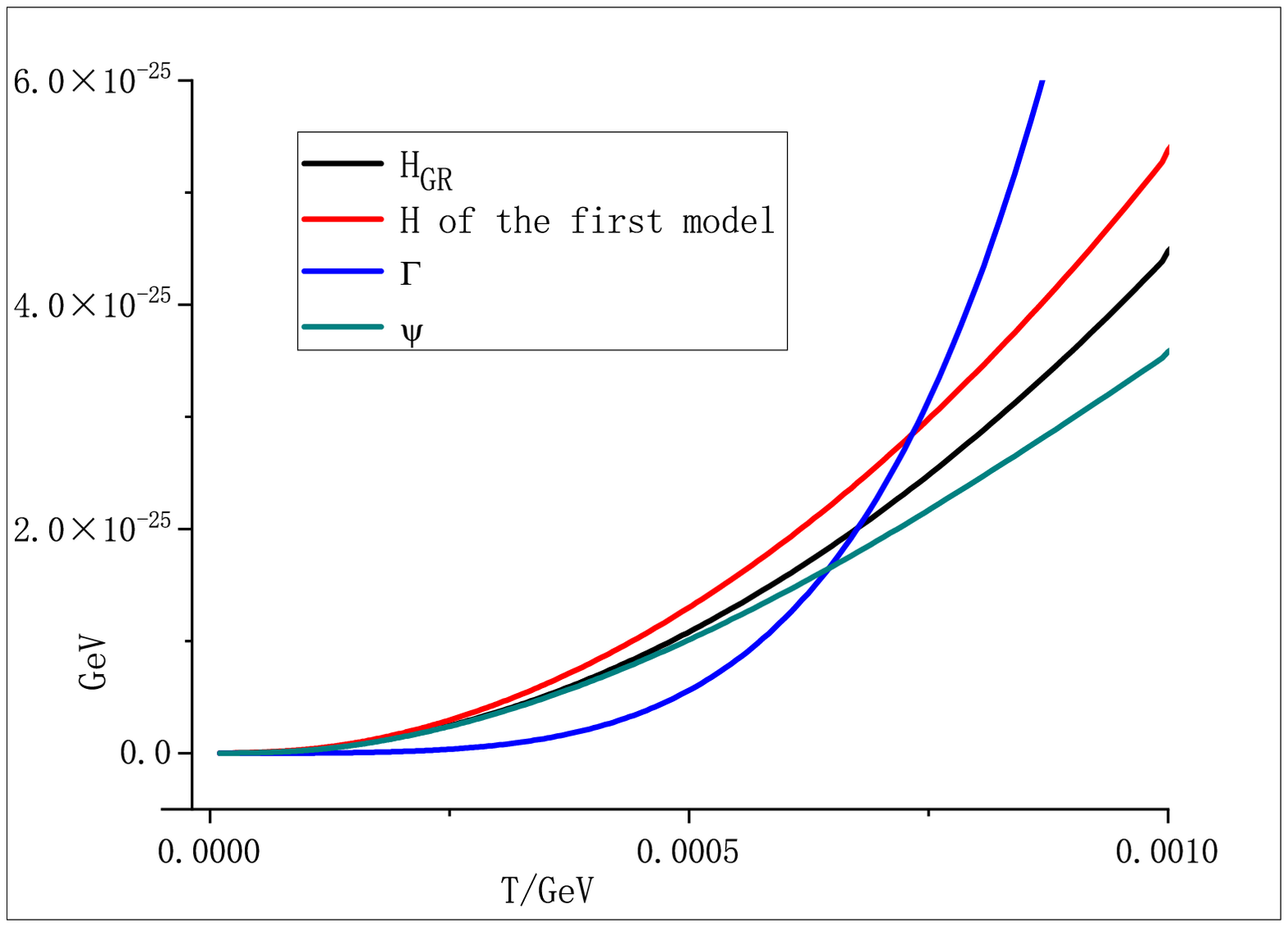}
 \includegraphics[scale=0.65]{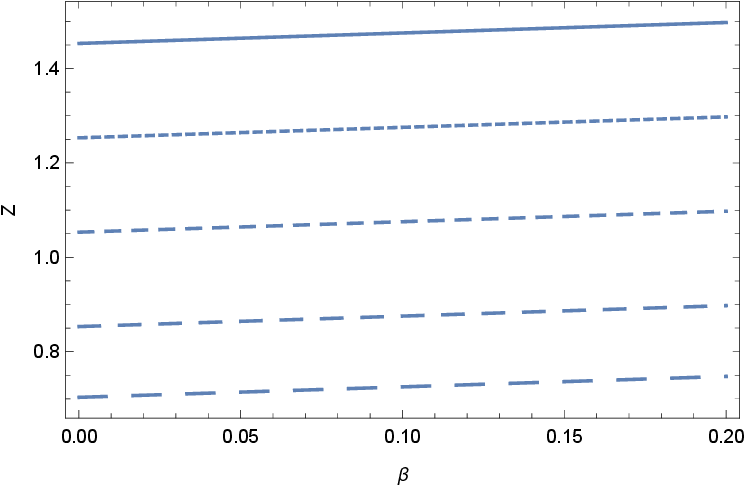}
\caption{Temperature evolution of $H_{GR}$, $\Gamma$,  $H$ and $\psi$ for the first Weyl type $f(Q,T)$  gravity model (left panel), and of the ratio $Z=H/H_{GR}$ (right panel). For the left panel plots the initial value of $\psi$, $\psi_{ini}$, is  $\psi_{ini}=0$, while the parameter values are given by $\alpha+\frac{m^{2}}{12\kappa^{2}}=-1$, and $\beta=1$, respectively. For the right panel plot $M=m/\sqrt{12}\kappa=0.08$, and the  values of $\alpha$ are $\alpha =-0.10$ (solid line), $\alpha =-0.50$ (dotted line), $\alpha =-0.90$ (short dashed line), $\alpha =-1.30$ (dashed line), and $\alpha =-1.60$ (long dashed line), respectively. }
\label{fig0}
\end{figure*}

\paragraph{The cosmological evolution in the radiation dominated era.} The comparative dynamical behavior of $H$, $H_{GR}$, $\psi$, $\Gamma$, and $Z$ is represented in Fig.~\ref{fig0}. As one can see from Fig.~\ref{fig0}, at hight temperatures there are significant differences between the Hubble functions of the $f(Q,T)$ model, and of the standard general relativity. These differences do decrease once the temperature of the Universe becomes smaller. The decay rate of the particles $\Gamma$ also decreases with the temperature, and at the freeze-out temperature it intersects the Hubble function. The Weyl vector $\psi$ also monotonically decreases with the temperature decrease.

\subsubsection{The freeze-out constraint.}

\begin{figure*}[ht]
\centering
\includegraphics[scale=0.3]{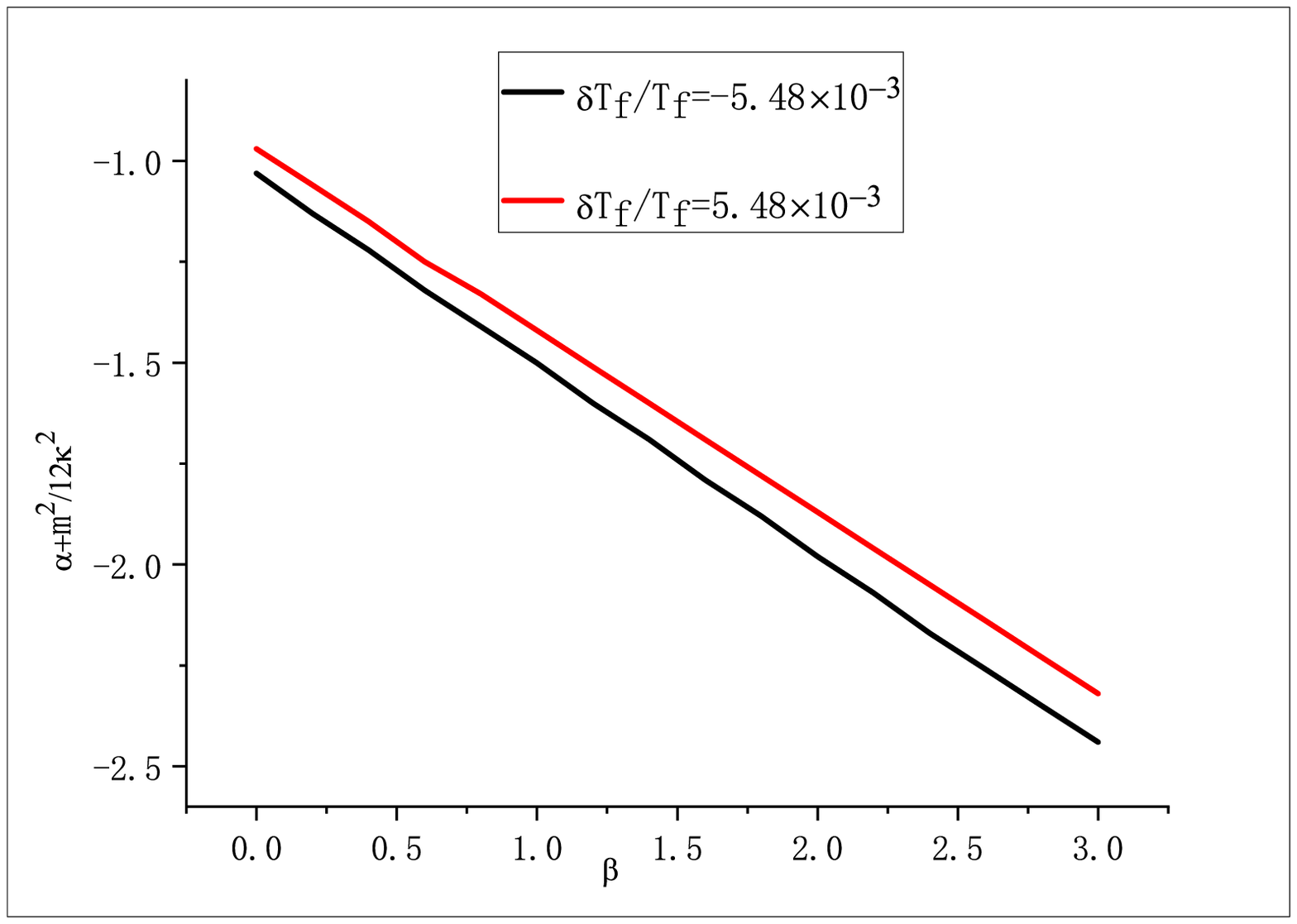}
 \includegraphics[scale=0.3]{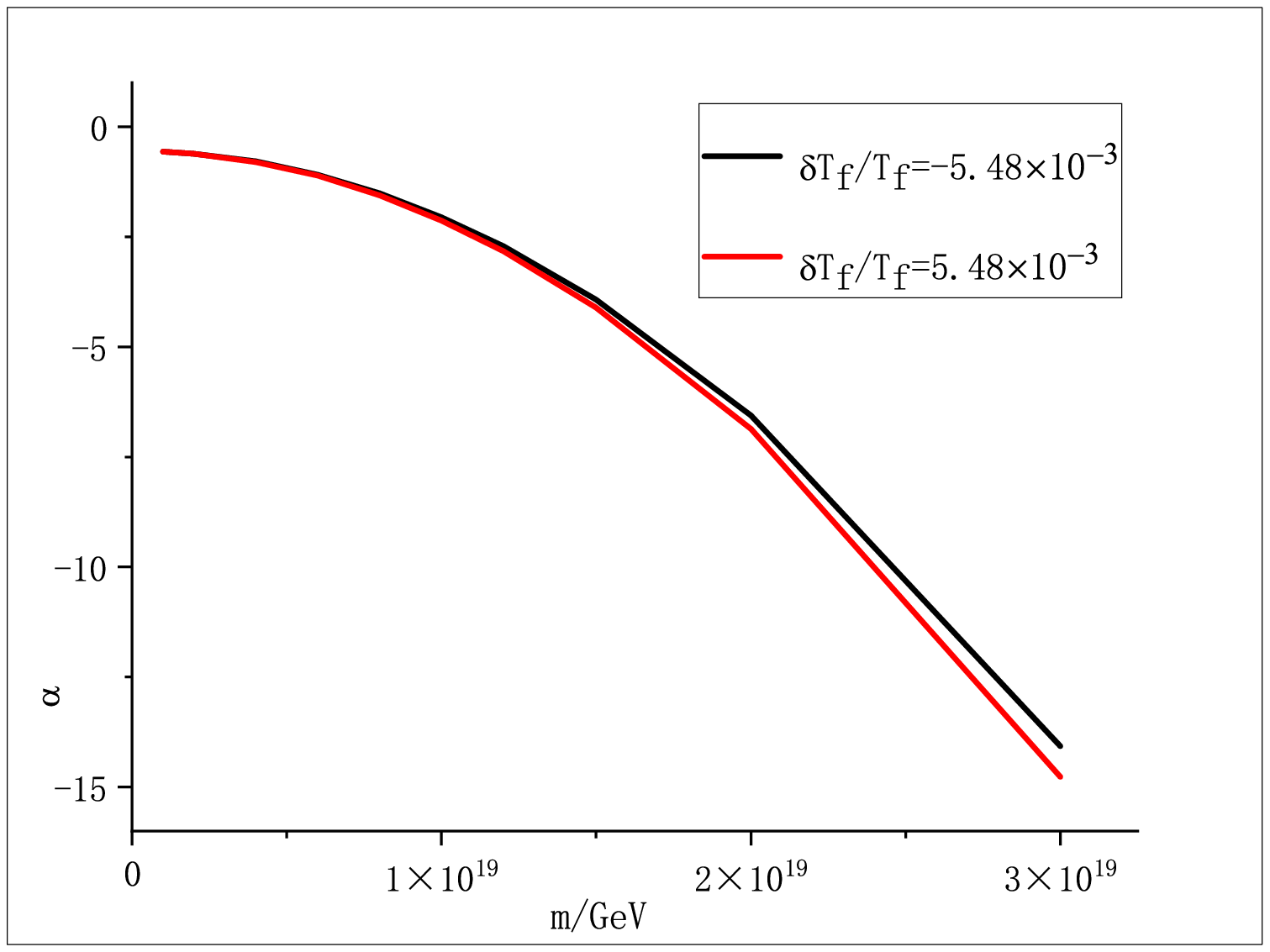}
\caption{Numerical freeze-out temperature constraint on the parameters $\left(\alpha+m^{2}/12\kappa^{2}, \beta\right)$ (left panel) and $(\alpha, m)$ (right panel) of the first Weyl type $f(Q,T)$ cosmological model. For the left panel plots  $\psi_{ini}=-1\times10^{-24}$ GeV, while for the right panel plots $\psi_{ini}=2\times10^{-25}$ GeV. The permitted region of the parameters is the area between the two curves. }
\label{fig01}
\end{figure*}

\paragraph{Numerical constraints.} As we have already mentioned, the freeze-out temperature is obtained as a solution of the equation $H\left(\mathcal{T}_f\right)=\Gamma \left(\mathcal{T}_f\right)$. The field equations can be integrated numerically for different values of the model parameter to obtain the variation of the Weyl vector $\psi$, and the freeze-out temperature $\mathcal{T}_f$. The numerical solution is obtained by adopting some initial values for $\psi$, which correspond to an initial temperature of $\mathcal{T}=2.7\times 10^{10}$ K. The constraints on the model parameters $\left(\alpha +m^2/12\kappa ^2,\beta\right)$ and $(\alpha,m)$, respectively, are presented in Fig.~\ref{fig01}.

\paragraph{The analytical freeze-out constraint.}The freeze-out constraint (\ref{53}) becomes in the first considered model of the Weyl type $f(Q,T)$ gravity
\be
\frac{\delta \mathcal{T}_f}{\mathcal{T}_f}=\frac{1}{10c_q}\sqrt{\frac{\pi^2g_{\ast}}{180\kappa ^2}}\frac{1}{\mathcal{T}_f^3}\Bigg(1+\alpha +\frac{4}{9}\beta +\frac{5\beta }{12\pi^2}\frac{m_\gamma ^2}{\mathcal{T}_f^2}+\frac{m^2}{12\kappa ^2}\Bigg).
\ee

By considering that the parameters $\alpha$ and $m^2/12\kappa ^2$ satisfy the condition $\alpha +m^2/12 \kappa ^2\approx -1$, and that the mass of the photon is negligible in the present context, for the parameter $\beta$ we obtain the constraint
\be\label{74}
\beta \approx \frac{45c_q}{2}\sqrt{\frac{180\kappa ^2}{\pi ^2g_\star}}\mathcal{T}^3_f\frac{\delta \mathcal{T}_f}{\mathcal{T}_f}.
\ee

For the numerical estimations we will consider  the following numerical values of the parameters: $g_\ast=10$, $\mathcal{T}_f=0.0006$ GeV, $c_q=9.8\times 10^{-10}$ GeV$^{-4}$, and $\lambda =\kappa ^2=M_p^2/16\pi$, where $M_p$, the Planck mass, is given by $M_p=1.22\times 10^{19}$ GeV.  Due to its smallness we will neglect the term related to the photon mass.

\begin{figure*}[ht]
\centering
\includegraphics[scale=0.50]{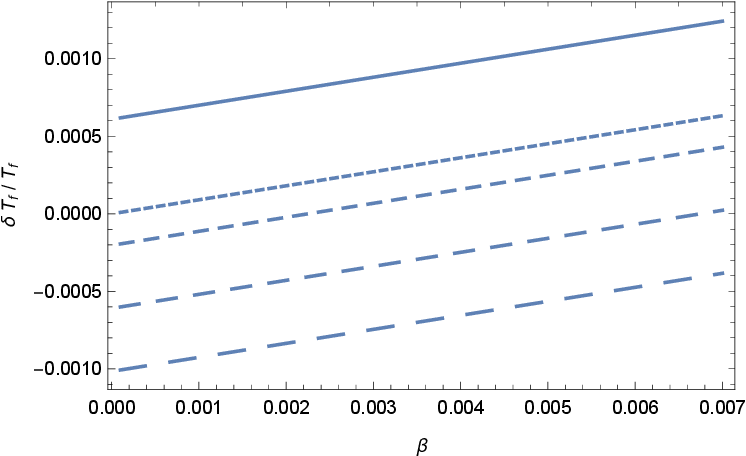}
\includegraphics[scale=0.50]{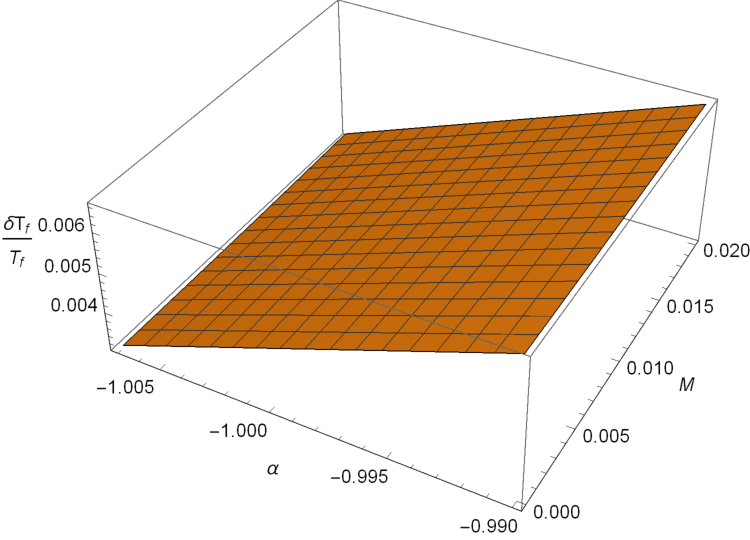}  %
\caption{ Variation of the ratio $\delta \mathcal{T}_f/\mathcal{T}_f$ as function of $\beta$ in the first Weyl type $f(Q,T)$ cosmological model (left panel), for $m^2/12\kappa ^2=10^{-8}$, and for different values of $\alpha$: $\alpha =-0.997$ (solid line), $\alpha =-1.000$ (dotted line), $\alpha =-1.001$ (short dashed line), $\alpha =-1.003$ (dashed line), and $\alpha =-1.005$ (long dashed line), respectively. In the right panel the variation of the ratio $\delta \mathcal{T}_f/\mathcal{T}_f$ is presented as function of $\alpha$ and $M=m/\sqrt{12}\kappa$ for $\beta =0.05$.}
\label{fig1}
\end{figure*}

By using the constraint (\ref{5}) it follows that $\beta <0.0005$. On the other hand, assuming $\alpha \approx -1$, it follows that the parameters $\beta $ and $m$ are given by
\be
\frac{4}{9}\beta +\frac{m^2}{12\kappa ^2}\approx 10c_q\sqrt{\frac{180\kappa ^2}{\pi ^2g_\star}}\mathcal{T}^3_f\frac{\delta \mathcal{T}_f}{\mathcal{T}_f}.
\ee
Hence we obtain the constraint $4\beta/9+m^2/12\kappa ^2<0.002$.
The variations of the ratio $\delta \mathcal{T}_f/\mathcal{T}_f$ are represented, for a fixed $\alpha$, and for different values of $\beta$, in the left panel of Fig.~\ref{fig1}, while the variation of $\delta \mathcal{T}_f/\mathcal{T}_f$ as a function of both $\alpha$ and $M=m/\sqrt{12}\kappa$ is represented in the right panel of the Figure.

% in Fig.~\ref{fig2}.

%\begin{figure}[ht]
%\centering
%\includegraphics[scale=0.60]{fig2.eps}  %
%\caption{ Variation of the ratio $\delta \mathcal{T}_f/\mathcal{T}_f$ as function of $\alpha$ and $M=m/\sqrt{12}\kappa$ for $\beta =0.05$.}
%\label{fig2}
%\end{figure}

\subsubsection{Deuterium and helium constraints.} Both the deuterium and helium constraints of the BBN can be satisfied if $Z$ is in the (approximate) range $Z\in(0.5, 1.6)$. The variation of $Z$ as a function of $\beta$, for fixed values of $\alpha$ and $M$, presented in the left panel of Fig.~\ref{fig0}, indicates that the present model satisfies both the deuterium and lithium constraints.

%\begin{figure}[ht]
%\centering
%\includegraphics[scale=0.60]{fig0.eps}  %
%\caption{ Variation of the ratio $Z=H/H_{GR}$ as function of $\beta $ for $M=m/\sqrt{12}\kappa=0.08$,  and for different values of $\alpha$: $\alpha =-0.10$ (solid line), %$\alpha =-0.50$ (dotted line), $\alpha =-0.90$ (short dashed line), $\alpha =-1.30$ (dashed line), and $\alpha =-1.60$ (long dashed line), respectively.}
%\label{fig3}
%\end{figure}

\subsubsection{The $^7 Li/H$ ratio.}

\begin{figure*}[ht]
\centering
\includegraphics[scale=0.3]{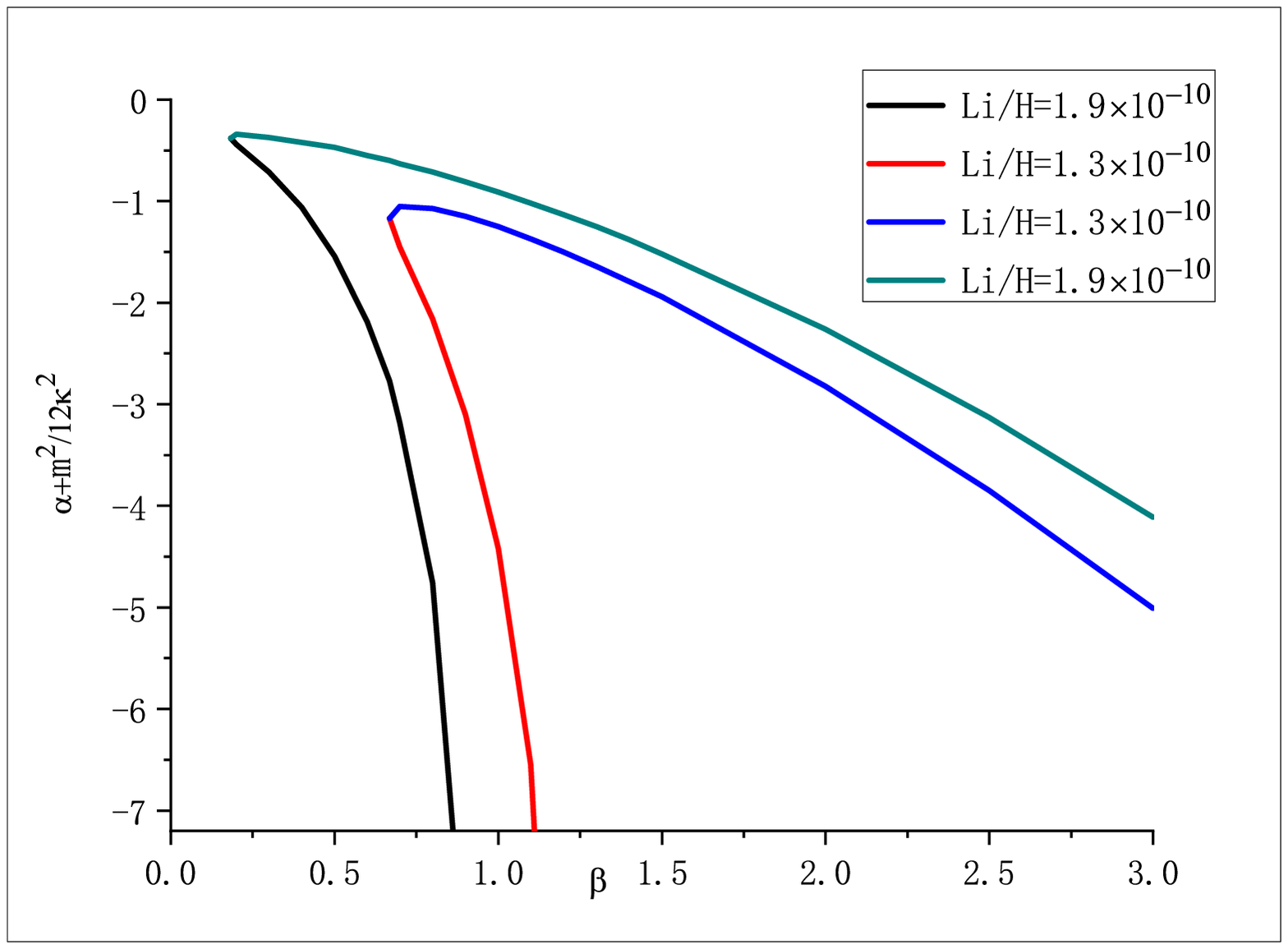}
 \includegraphics[scale=0.3]{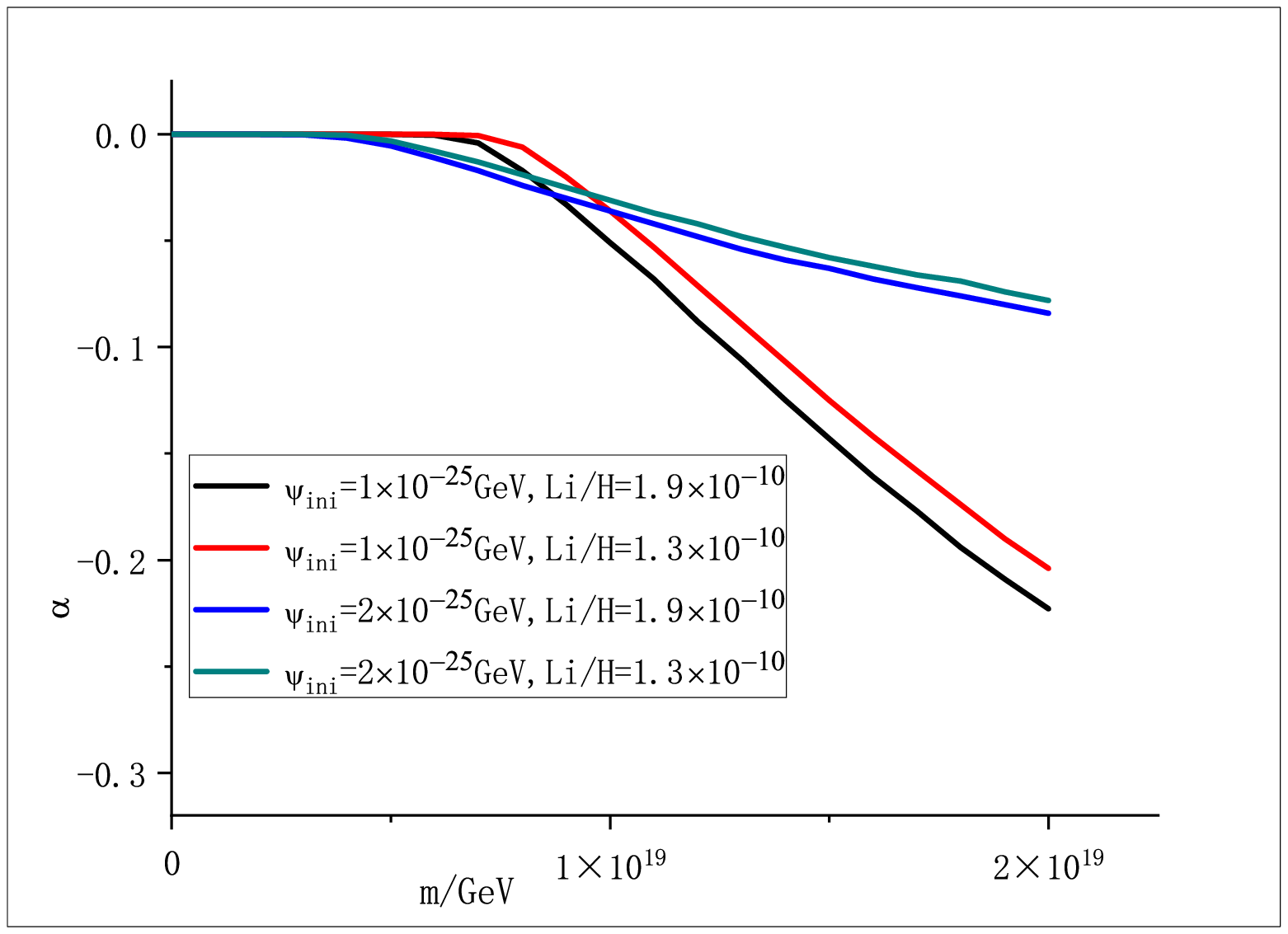}
\caption{Numerical Lithium abundance constraints on the parameters $\left(\alpha+m^{2}/12\kappa^{2}, \beta\right)$ (left panel) and $(\alpha,m)$ (right panel)
 of the first Weyl type $f(Q,T)$ cosmological model for different initial values of the Weyl vector $\psi$. }
\label{fig02}
\end{figure*}

\paragraph{Numerical results.} In this Subsubsection, we will combine the $^{4}He$ constraints and the $^{7}Li$ constrains to obtain the range of the model parameters. The results obtained through the numerical integration of the generalized Friedmann equations are presented in Fig.~\ref{fig02}.

From the numerical analysis one obtains the constraints on the model parameters as
\begin{equation}
    -1.35<\alpha+\frac{m^{2}}{12\kappa^{2}}<-1.16,\;0.43<\beta<0.69.
\end{equation}
and
\begin{equation}
    -1.83<\alpha+\frac{m^{2}}{12\kappa^{2}}<-1.53,\; 1.22<\beta<1.70,
\end{equation}
respectively.

\paragraph{Analytical results.} In the first considered cosmological model in the Weyl type $f(Q,T)$ gravity, the Lithium abundance is obtained as
\bea
\frac{^{7}Li}{H}&=&4.82\times 10^{-10}\times (1\pm 0.1)\nonumber\\
&&\times \left[ \frac{%
10^{10}\eta -(3/2)\left( 1+\alpha +4\beta /9+m^{2}/12\kappa ^{2}\right) }{6}%
\right] ^{2}.\nonumber\\
\eea

\begin{figure}[ht]
\centering
\includegraphics[scale=0.60]{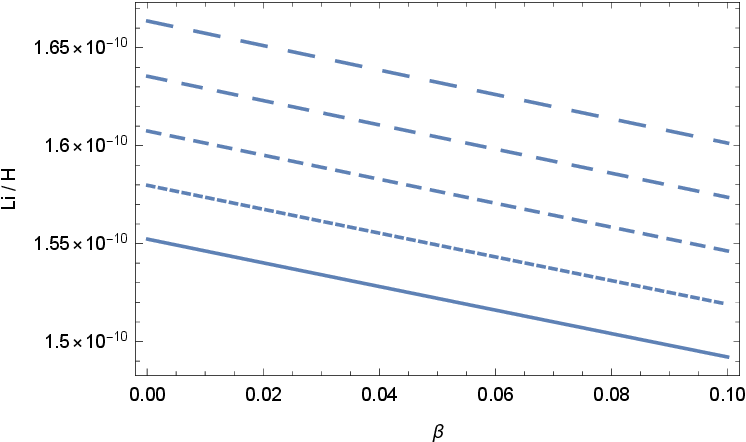}  %
\caption{ Variation of the ratio $^7 Li/H$ as function of $\beta$ for $M=m/\sqrt{12}\kappa=0.10$, and for different values of $\alpha$:   $\alpha =0.72$ (solid line), $\alpha =0.70$ (dotted line), $\alpha =0.68$ (short dashed line), $\alpha =0.66$ (dashed line), and $\alpha =0.64$ (long dashed line), respectively.}
\label{fig4a}
\end{figure}

The variation of the  $^7 Li/H$ ratio as a function of the model parameter $\beta$, for fixed $m^2/12\kappa ^2$ values, different values of $\alpha$, is represented in Fig.~\ref{fig4a}.

\begin{figure*}[ht]
\centering
\includegraphics[scale=0.25]{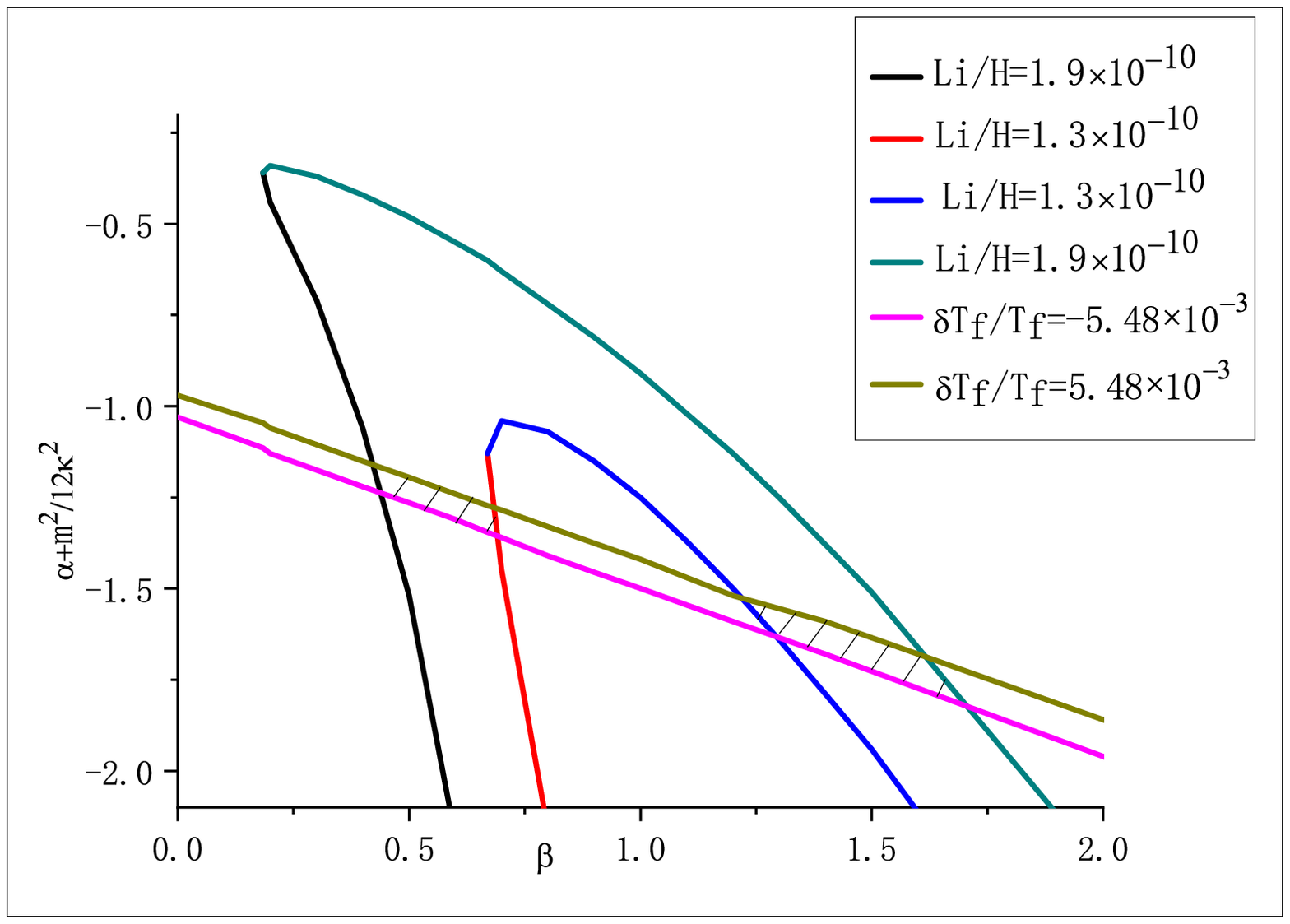}
\includegraphics[scale=0.25]{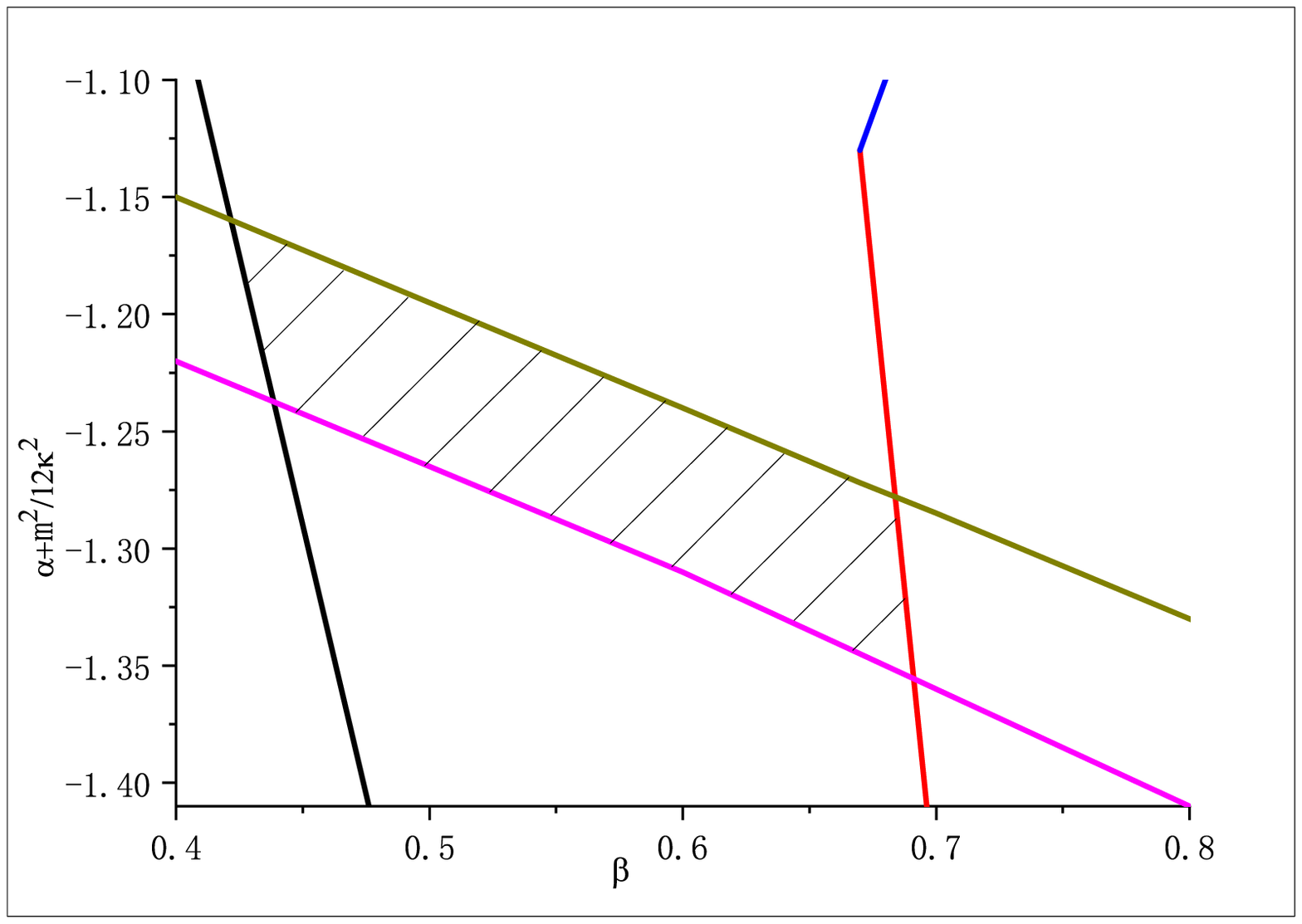}%
\includegraphics[scale=0.25]{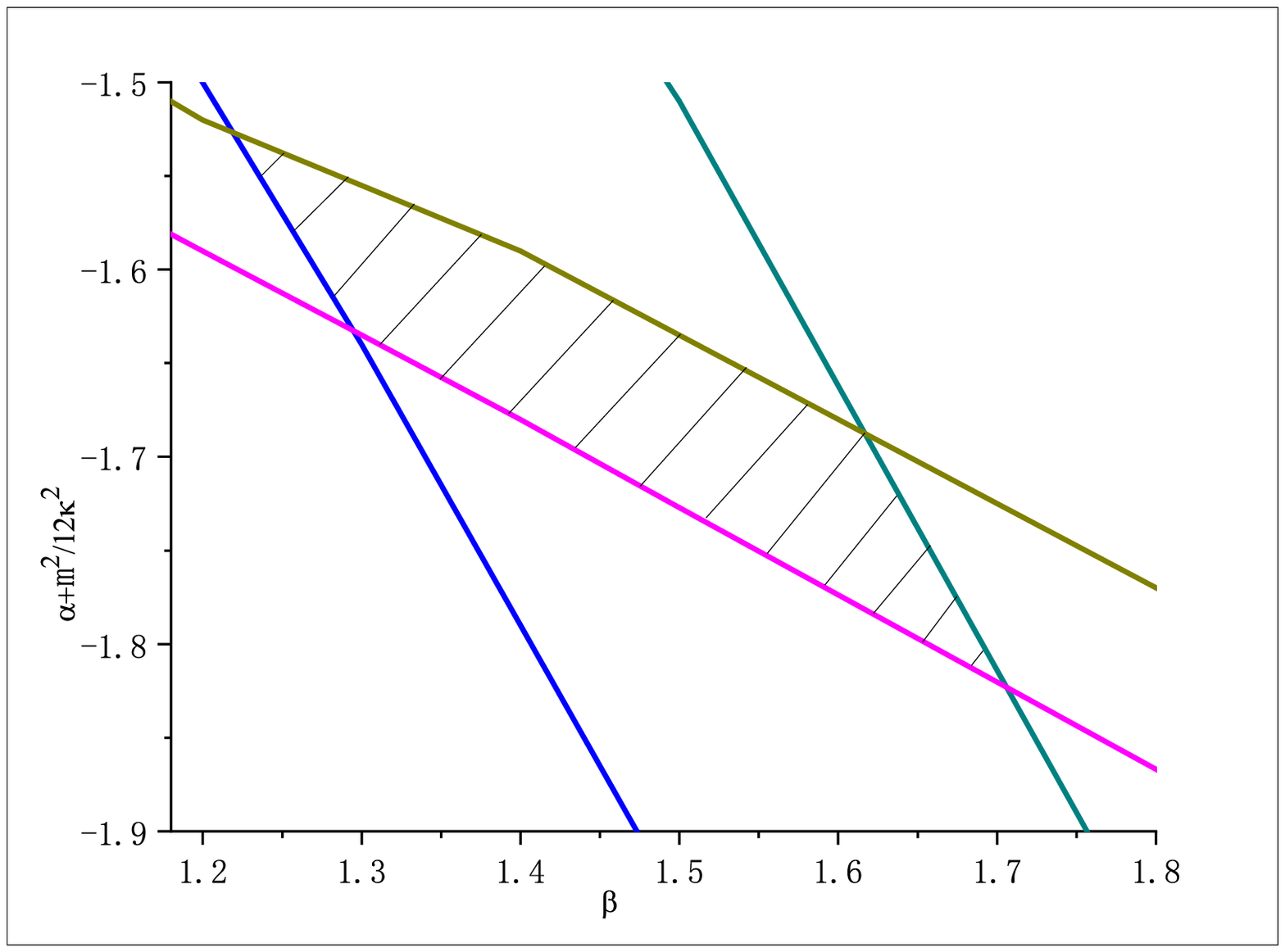}
\caption{The combined  freeze-out temperature and Lithium abundance constraint on the first model parameters $\left(\alpha +m^2/12\kappa ^2\right)$ and $\beta$. For the initial values of $\psi$ we have considered the values $-1\times 10^{-24}$ GeV, $0$, and $1\times 10^{-24}$ GeV. The allowed range of parameters lies in the shadowed area. }
\label{fig5n}
\end{figure*}

\paragraph{The combined $^7$Li and freeze-out constraint.} The combined freeze-out temperature and $^7$Li constraints are represented in Fig.~\ref{fig5n}.

\subsubsection{Model II: $f(Q,T)=\frac{\gamma}{6H_{0}^{2}\kappa^{2}}QT$}

As a second example of an $f(Q,T)$ type cosmological model we consider the case in which the function $f$ is given by
\begin{equation}
    f(Q,T)=\frac{\gamma}{6H_{0}^{2}\kappa^{2}}QT,
\end{equation}
where $\gamma $ is a constant, and $H_0$, the Hubble rate at the freeze-out point, is given by $H_0=H_{GR}\left(\mathcal{T}_f\right)=2.42\times 10^{-25}$ GeV.
For this model, we have two modified Friedmann equations
\bea
H^{2}&=&\frac{1}{6\lambda}\Bigg[\rho+(m^{2}+12\lambda)(H\psi-\frac{1}{2}\psi^{2})+\frac{\gamma}{H_{0}^{2}}(6p-2\rho)H\psi\nonumber\\
        &&+\frac{\gamma}{H_{0}^{2}}(3\rho-p)\psi^{2}\Bigg],
\eea
\bea
\dot{H}&=&\frac{-1}{4\lambda}\Bigg[(1+\frac{2\gamma \psi^{2}}{H_{0}^{2}})(\rho+p)-\frac{1}{3}(m^{2}+\frac{6\gamma }{H_{0}^{2}}p-\frac{2\gamma }{H_{0}^{2}}\rho\nonumber\\
        &&+12\lambda)(\dot{\psi}+\psi^{2}-H\psi)-\frac{2\gamma}{3H_{0}^{2}}(3\dot{p}-\dot{\rho})\psi\Bigg].
\eea

In the following we adopt again the approximation $\lambda =\kappa ^2$.

\begin{figure*}[ht]
\centering
\includegraphics[scale=0.3]{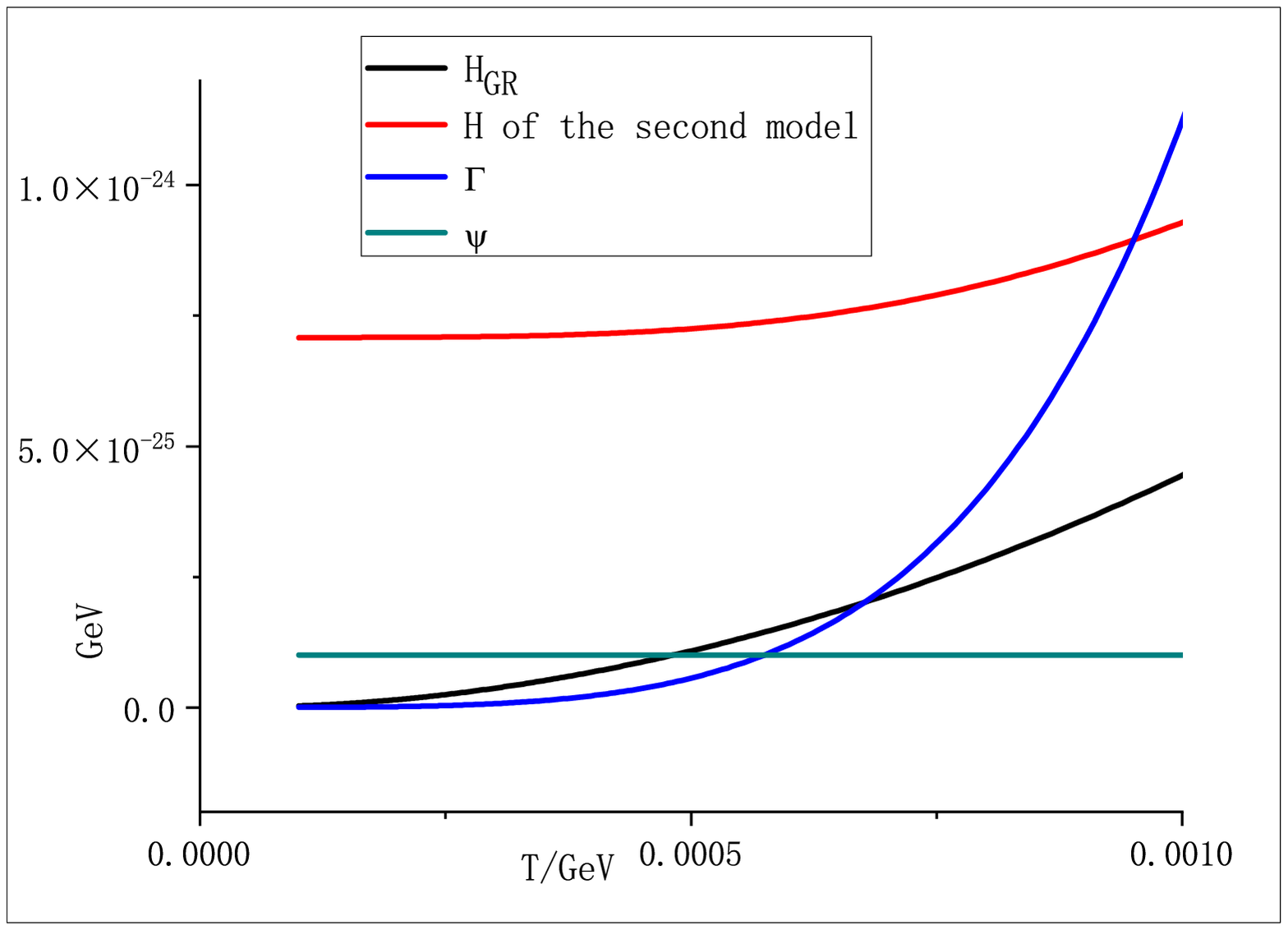}
 \includegraphics[scale=0.3]{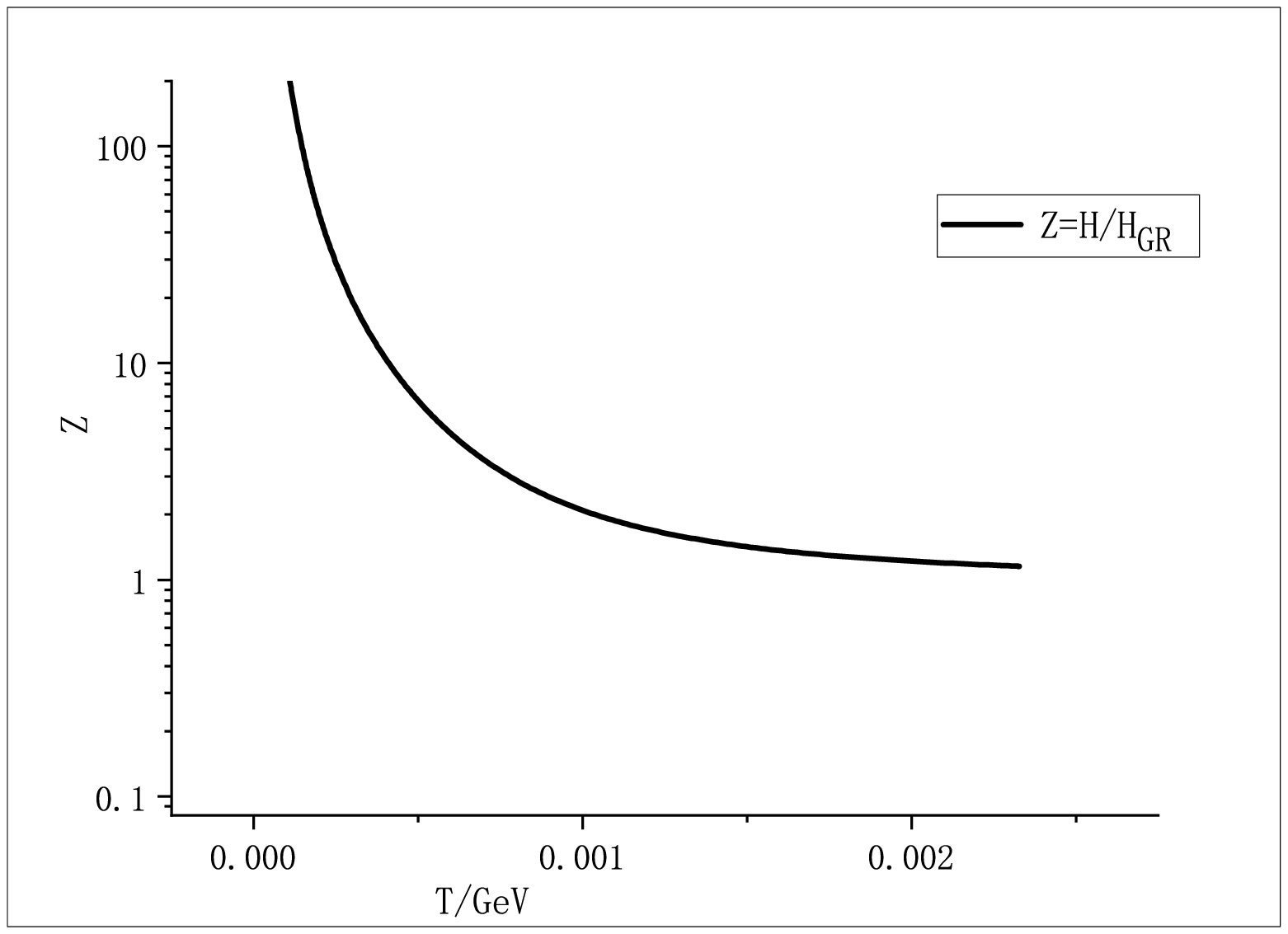}
\caption{Temperature evolution of $H_{GR}$, $\Gamma$, $H$ and $\psi$ for the second Weyl type $f(Q,T)$ cosmological model (left panel), and of the ratio $Z=H/H_{GR}$ (right panel). For the left panel plots the initial value of $\psi$, $\psi_{ini}$, is  $\psi_{ini}=1\times 10^{-25}$ GeV, while $\gamma =-1$. For the right panel plot $\psi_{ini}=10^{-25}$ GeV.  }
\label{fig4}
\end{figure*}

\paragraph{Cosmological evolution in the radiation dominated epoch.} The variations of the functions $H$, $H_{GR}$, $\psi$, $\Gamma$, and of $Z$ are represented, for the second model, in Fig.~\ref{fig4}. In this case there are significant differences between the evolution of the general relativistic Hubble function, and the Hubble function of the model, with the differences increasing with decreasing temperature. This behavior can be also observed from the variation of $Z$, indicating significant differences, of two orders of magnitude, at low temperatures. However, for enough high temperatures of the order of 0.025 GeV, and higher, the two Hubble functions basically coincide. For temperatures lower than 0.0005 GeV, the Hubble function $H$ becomes approximately a constant, indicating the transition to an accelerated, de Sitter type evolution. During the entire period of the cosmological evolution in the radiation dominated phase, the Weyl vector is a constant, and thus it plays the role of an effective cosmological constant. The freeze-out temperature, obtained from the intersection of the decay rate $\Gamma$ and the Hubble functions, takes very different values in the two models.

For the function $\rho_{eff}/\rho_r$ we obtain
\be
\frac{\rho_{eff}}{\rho_r}=1-\frac{\gamma g_\ast m_\gamma ^2}{72H_0^2\kappa ^2}\mathcal{T}^2+\frac{2\pi^2\gamma  g_\ast}{135 H_0^2\kappa ^2}\mathcal{T}^4+\frac{m^2}{12 \kappa ^2}.
\ee

\subsubsection{The freeze-out constraint}

\begin{figure*}[ht]
\centering
\includegraphics[scale=0.3]{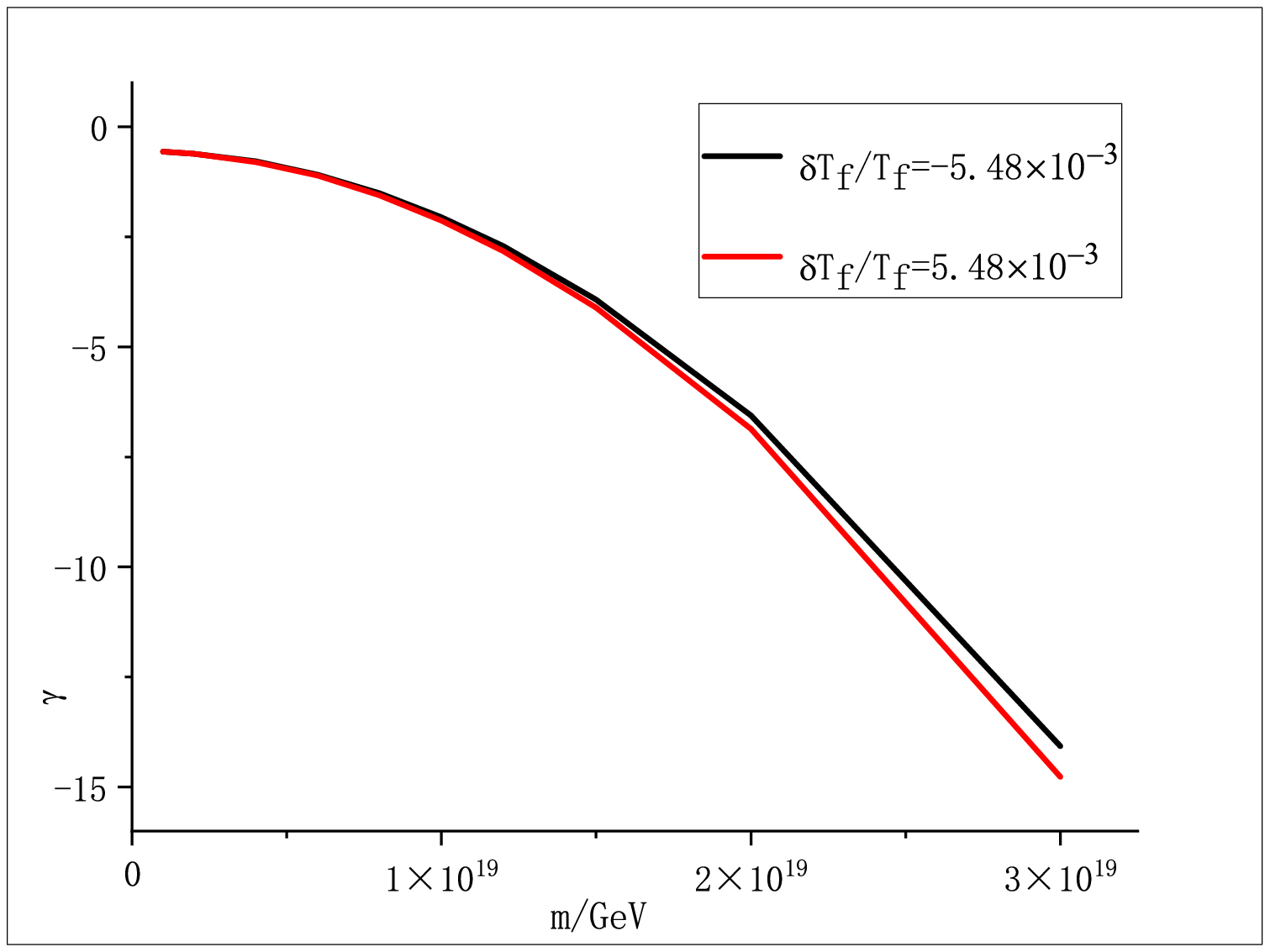}
 \includegraphics[scale=0.3]{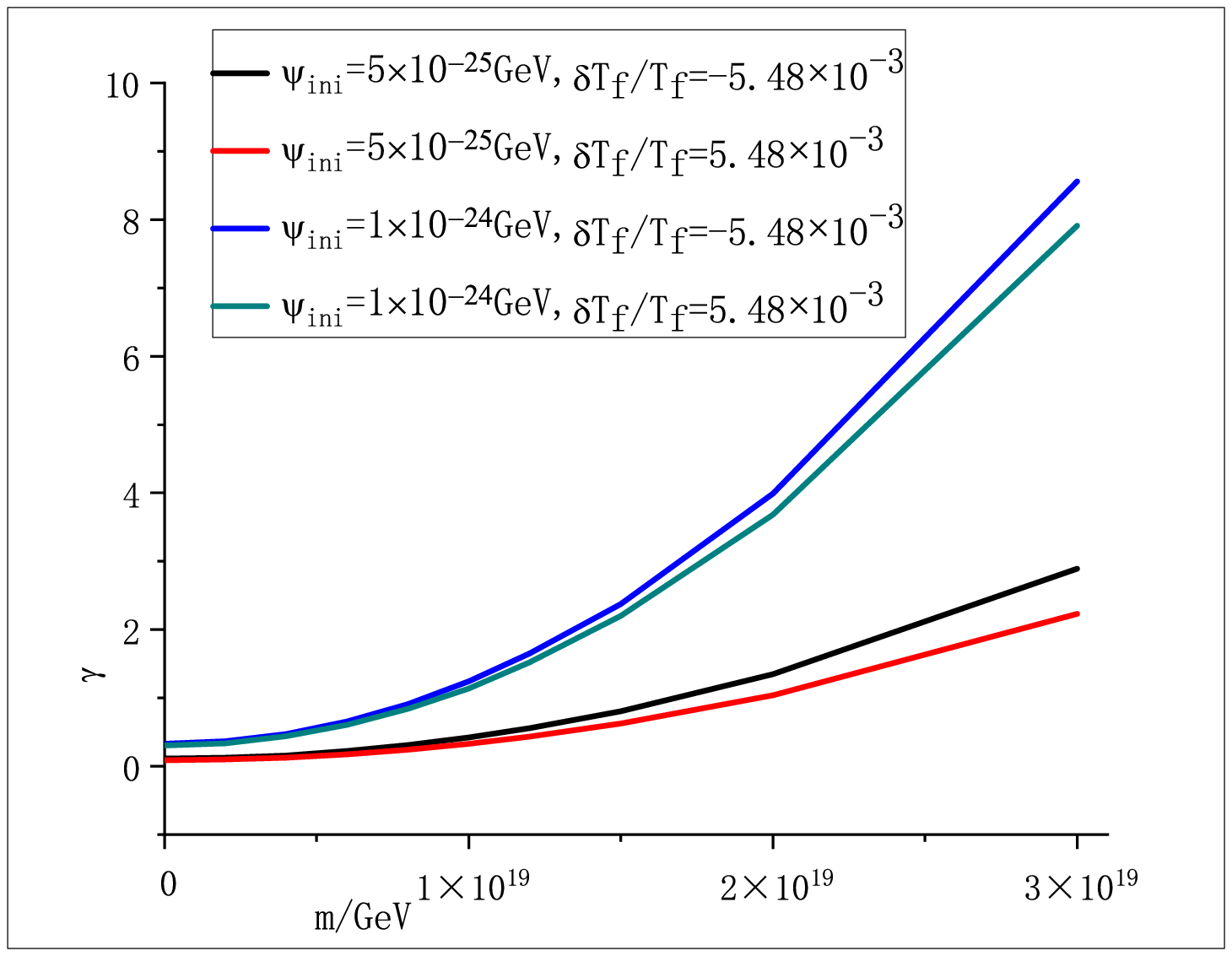}
\caption{The freeze-out constraint on the parameters $\gamma$ and $m$ of the second Weyl type cosmological model. For the left panel curves the initial values of the Weyl vector are $1\times 10^{-25}$ GeV, and $2\times 10^{-25}$ GeV, respectively, while the right panel curves have been obtained for the initial values of $\psi$ given by $5\times 10^{-25}$ GeV, and $1\times 10^{-24}$ GeV, respectively.}
\label{fig2Tf}
\end{figure*}

\paragraph{The numerical freeze-out constraint.}The freeze-out constraint on the parameters of the second model, obtained by numerically solving the gravitational field equations for different initial values of the Weyl vector $\psi$,  are represented in Fig.~\ref{fig2Tf}, for both positive and negative values of $\gamma$. As one can see from the Figure, the behavior of the solution of the constraint equation $H\left(\mathcal{T}_f\right)=\Gamma \left(\mathcal{T}_f\right)$ is strongly dependent on the sign, and on the numerical value of $\gamma$ and $m$. However, for both signs of $]\gamma$ one can obtain a range of numerical values of $(\gamma, m)$ satisfying the observational freeze-out constraint.

\paragraph{Analytical results.} For the second Weyl type $f(Q,T)$ theory cosmological model the freeze-out constraint (\ref{53}) takes the form
\begin{eqnarray}
\left\vert \frac{\delta \mathcal{T}_{f}}{\mathcal{T}_{f}}\right\vert  &=&%
\frac{1}{10c_{q}}\sqrt{\frac{\pi ^{2}g_{\ast }}{180\kappa ^{2}}}\frac{1}{%
\mathcal{T}_{f}^{3}}\Bigg(1+\frac{\gamma }{6H_{0}^{2}\kappa ^{2}}T  \nonumber\\
&&+\frac{4\pi ^{2}g_{\ast }\mathcal{T}%
_{f}^{4}}{270H_{0}^{2}\kappa ^{2}}\gamma +\frac{m^{2}}{12\kappa ^{2}}\Bigg).
\end{eqnarray}

By neglecting the trace of the matter energy-momentum tensor, we obtain the freeze-out temperature BBN constraint as
\begin{equation}
1+\frac{4\pi ^{2}g_{\ast }\mathcal{T}_{f}^{4}}{270H_{0}^{2}\kappa ^{2}}%
\gamma +\frac{m^{2}}{12\kappa ^{2}}<0.002.
\end{equation}

By taking into account that $\left( 4\pi ^{2}g_{\ast }\mathcal{T}%
_{f}^{4}/270H_{0}^{2}\kappa ^{2}\right) =1.09$, we obtain the upper limit of
the model parameters as given by

\begin{equation}
1+1.09\gamma +\frac{m^{2}}{12\kappa ^{2}}<0.002.
\end{equation}

This constraint is satisfied only for negative values of $\gamma $.

%\subsubsection{Deuterium and Helium constraints}

\subsubsection{The combined $^7 Li/H$ ratio}

\begin{figure*}[ht]
\centering
\includegraphics[scale=0.3]{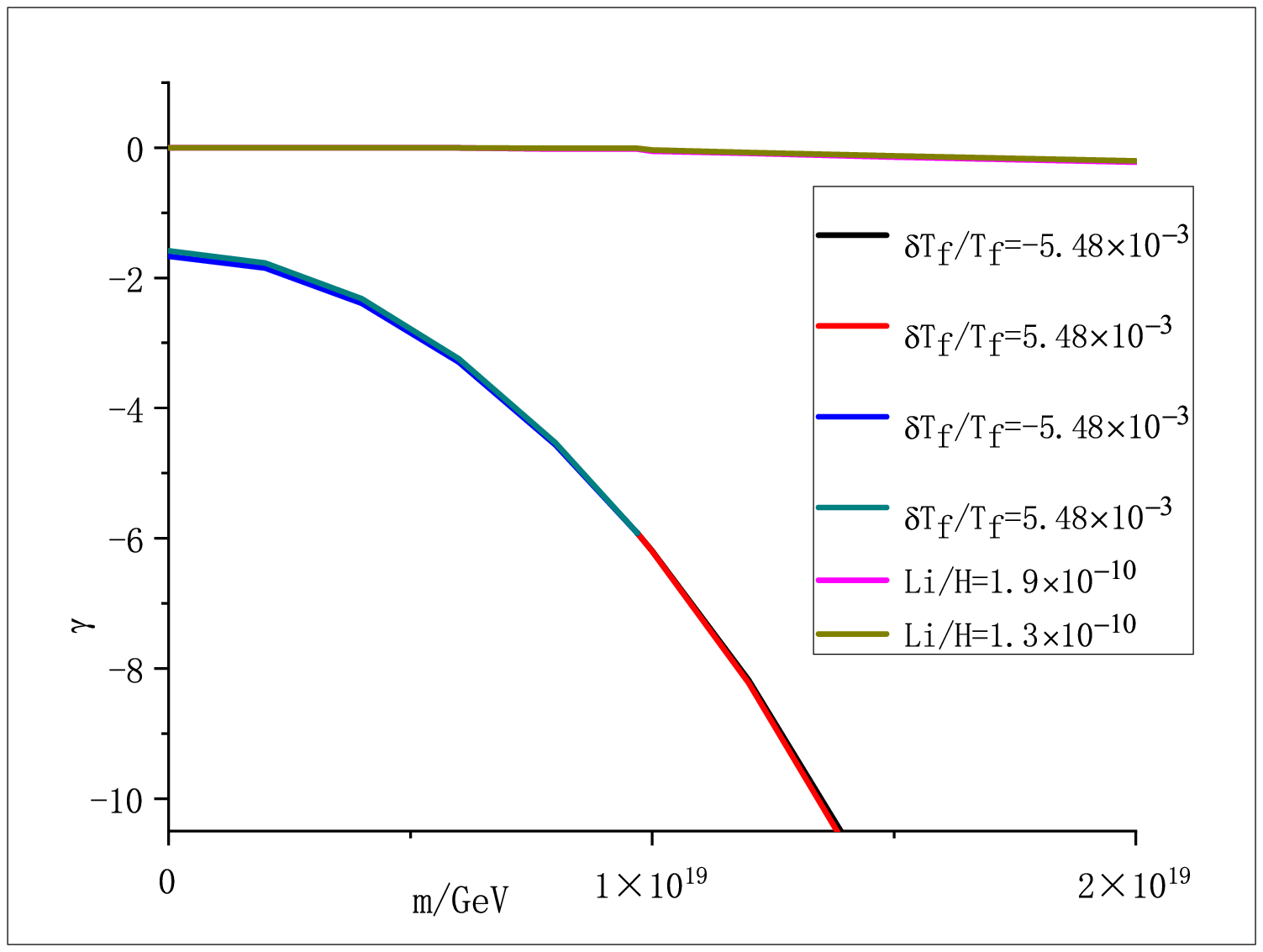}
\includegraphics[scale=0.3]{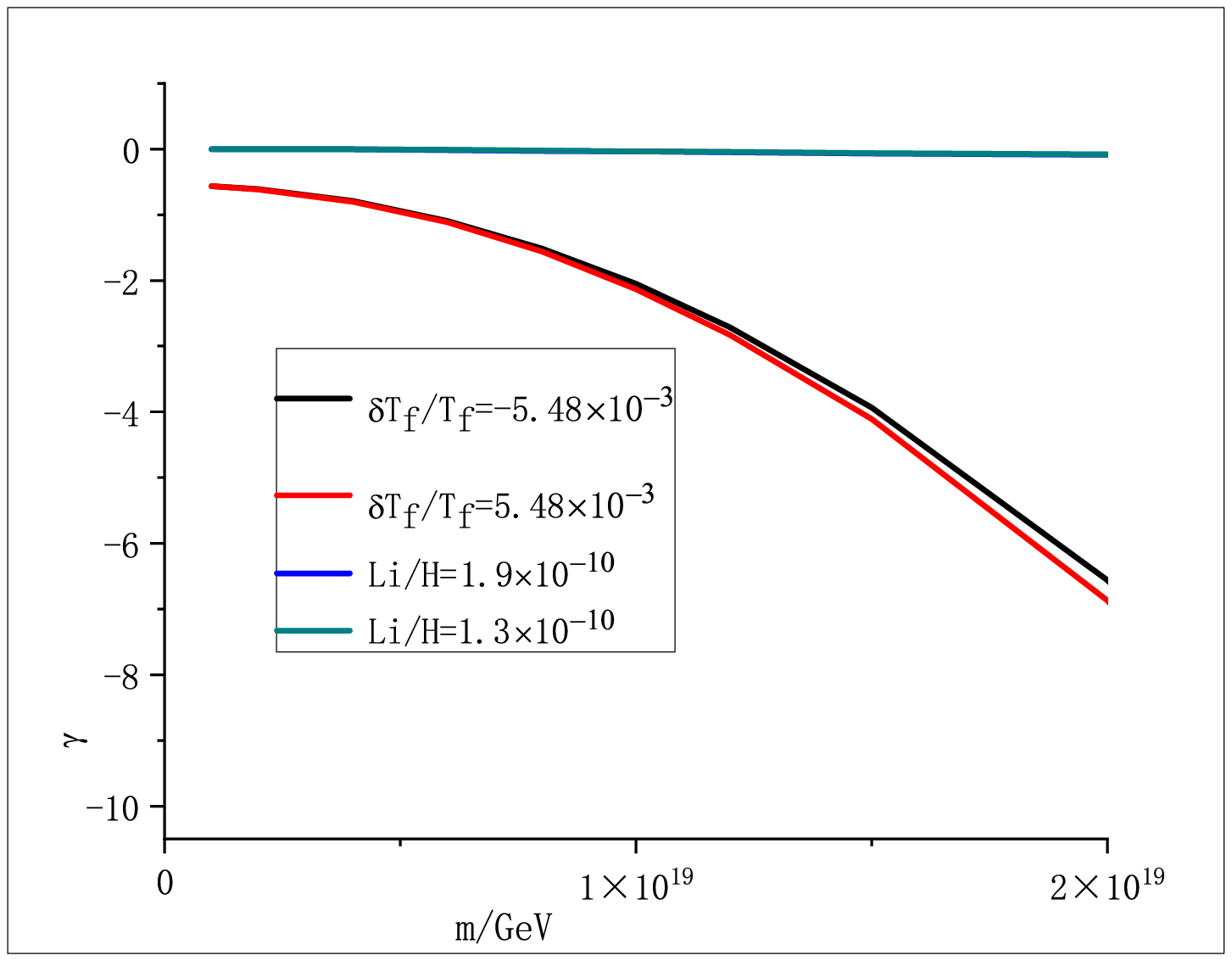}
\includegraphics[scale=0.3]{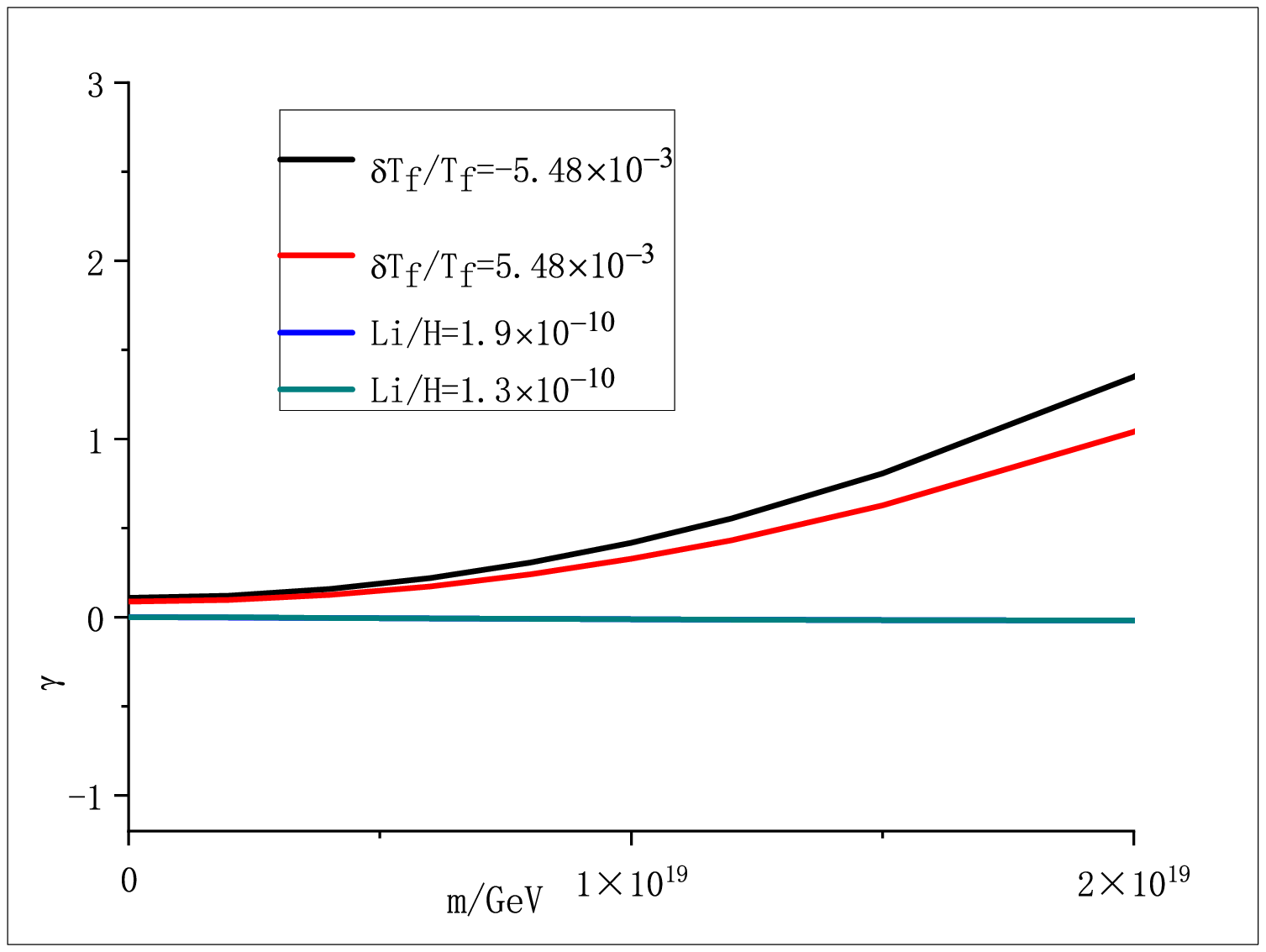}
 \includegraphics[scale=0.3]{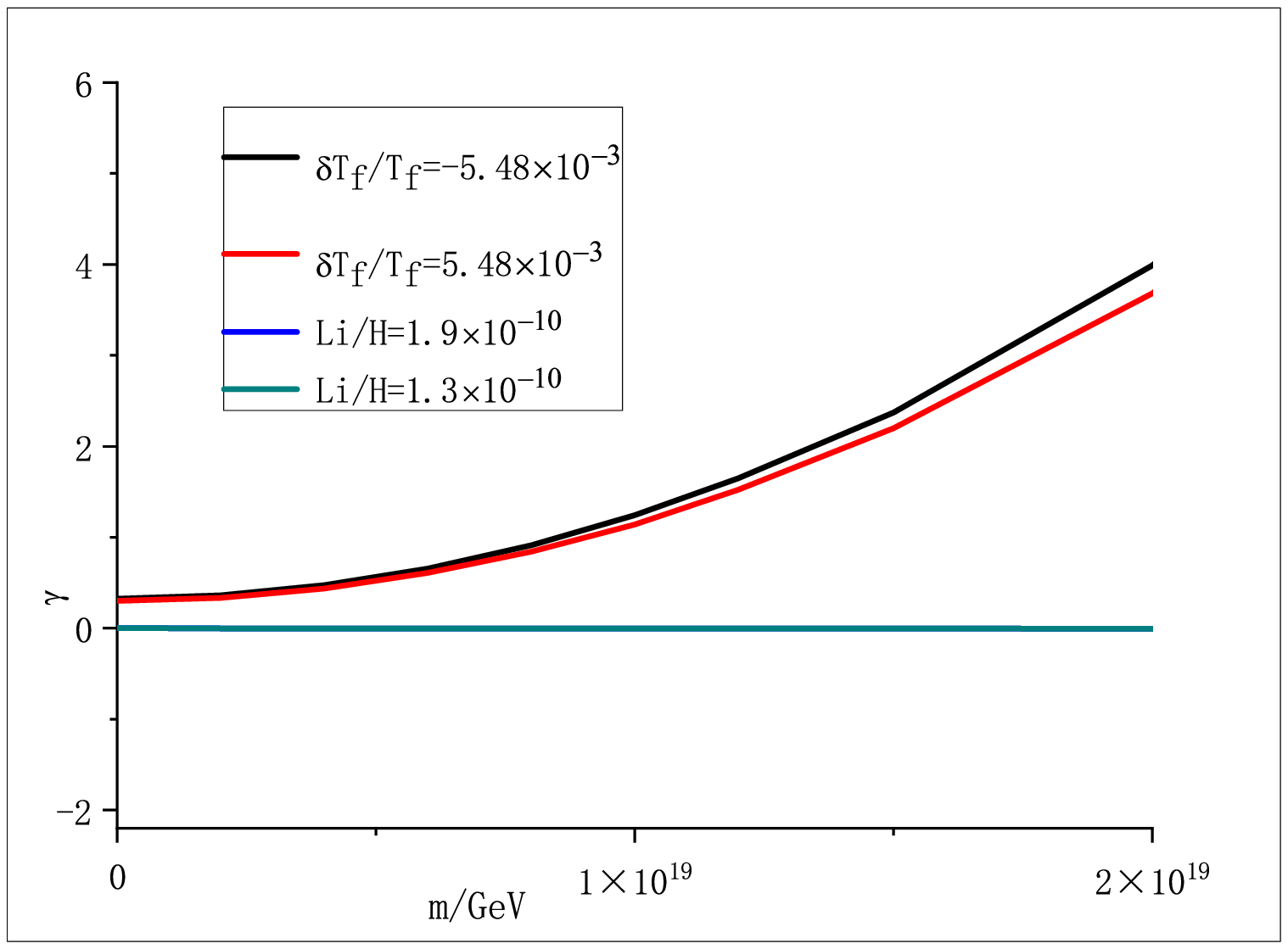}
\caption{The combined freeze-out temperature and Lithium constraints on the parameters $\gamma$ and $m$ of the second Weyl type cosmological model. The initial values of the Weyl vector are $1\times 10^{-25}$ GeV,  $2\times 10^{-25}$ GeV,  $5\times 10^{-25}$ GeV, and $1\times 10^{-24}$ GeV, respectively.}
\label{fig2Li}
\end{figure*}

For the second Weyl type $f(Q,T)$ modified gravity theory the combined Lithium and freeze-out temperature constraints on the model parameters $\gamma$ and $m$ are depicted in Fig.~\ref{fig2Li}.

\subsection{Model III: $f(Q,T)=\eta H_{0}^{2}e^{\frac{\mu}{6H_{0}^{2}}Q}+\frac{\nu}{6\kappa^{2}}T$}

As a third example of Weyl type $f(Q,T)$ models we assume that the function $f$ can be written as
\begin{equation}
    f(Q,T)=\eta H_{0}^{2}e^{\frac{\mu}{6H_{0}^{2}}Q}+\frac{\nu}{6\kappa^{2}}T,
\end{equation}
where $\mu$, $\nu$ and $\eta$ are constants. The two modified Friedmann equations of this Weyl type $f(Q,T)$ model can be written as
\bea
\hspace{-1.0cm}H^{2}&=&\frac{1}{6\lambda}\Bigg[\rho+\frac{\nu}{6}(3\rho-p)-\kappa^{2}\eta H_{0}^{2}e^{\frac{\mu}{H_{0}^{2}}\psi^{2}}+\Big(m^{2}\nonumber\\
         &&+2\kappa^{2}\eta\mu e^{\frac{\mu}{H_{0}^{2}}\psi^{2}}+12\lambda\Big)H\psi-\frac{m^{2}\psi^{2}}{2}-6\lambda\psi^{2}\Bigg],
\eea
\bea
\hspace{-1.0cm}\dot{H}&=&\frac{-1}{4\lambda}\Bigg[\left(1+\frac{\nu}{3}\right)(\rho+p)-\frac{1}{3}\Big(m^{2}+2\kappa^{2}\eta\mu e^{\frac{\mu}{H_{0}^{2}}\psi^{2}}\nonumber\\
        &&+12\lambda\Big)\left(\dot{\psi}+\psi^{2}-H\psi\right)-\frac{4\kappa^{2}\eta\mu^{2}\psi^{2}\dot{\psi}}{3H_{0}^{2}}e^{\frac{\mu}{H_{0}^{2}}\psi^{2}}\Bigg].\nonumber\\
\eea
We again adopt the approximation $\lambda =\kappa ^2$.

\begin{figure*}[ht]
\centering
\includegraphics[scale=0.3]{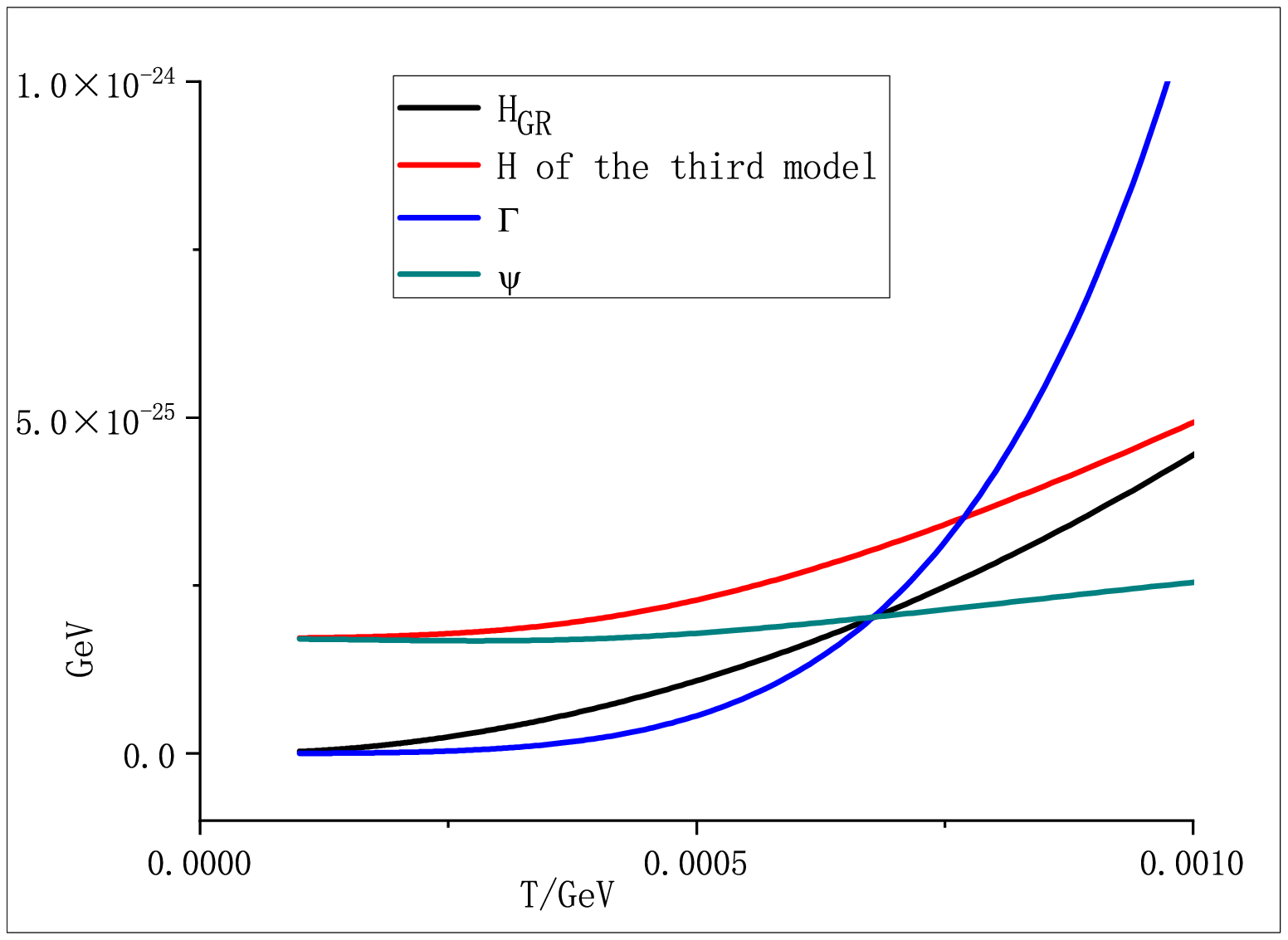}
 \includegraphics[scale=0.3]{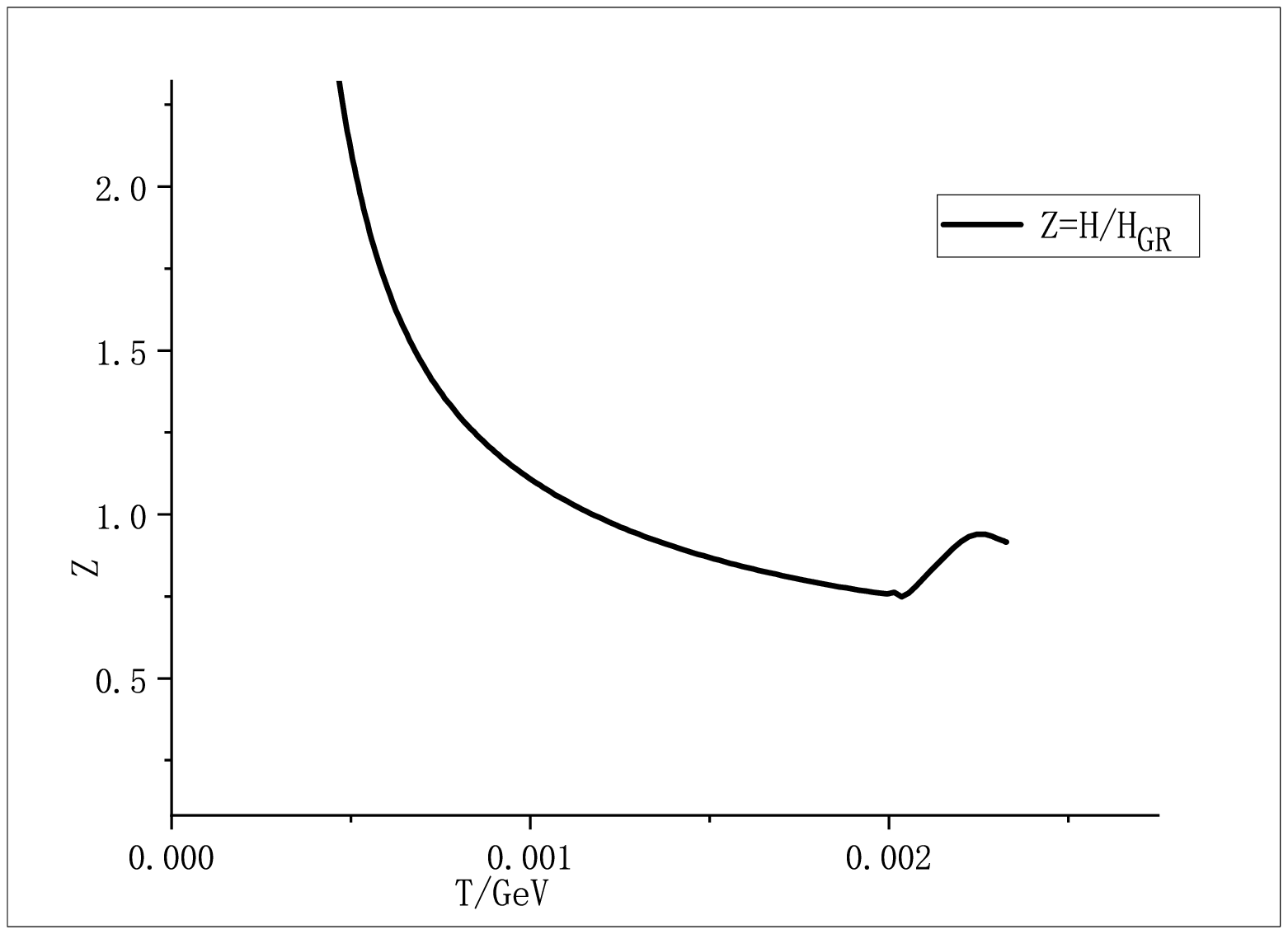}
\caption{Temperature evolution of $H_{GR}$, $\Gamma$, $H$, and $\psi$ for the third Weyl type $f(Q,T)$ cosmological model (left panel), and of the ratio $Z=H/H_{GR}$ (right panel). For the left panel plots the initial value of $\psi$, $\psi_{ini}$, is  $\psi_{ini}=1\times 10^{-25}$ GeV, while $\gamma =-1$. For the right panel plot $\psi_{ini}=10^{-25}$ GeV. }
\label{fig5}
\end{figure*}

\paragraph{The cosmological evolution.} For the third Weyl type $f(Q,T)$ cosmological model the comparative variations of the functions $H$, $H_{GR}$, $\psi$, $\Gamma$, and $Z$ are represented in Fig.~\ref{fig5}. At enough high temperatures the behaviors of the two Hubble functions become similar, but significant differences do exist in the lower temperature regime, during the BBN phase. However, at the freeze-out time, $Z$ takes values of the order of two, or smaller. The Weyl vector decreases from the high temperature regime up to around the BBN epoch, and then  it becomes a constant, indicating again the possible transition to an accelerated phase in the evolution of the Universe. But once the transition to the matter dominated era takes place,  the Universe may enter into a decelerating phase.

\subsubsection{The freeze-out constraint}

\begin{figure*}[ht]
\centering
\includegraphics[scale=0.3]{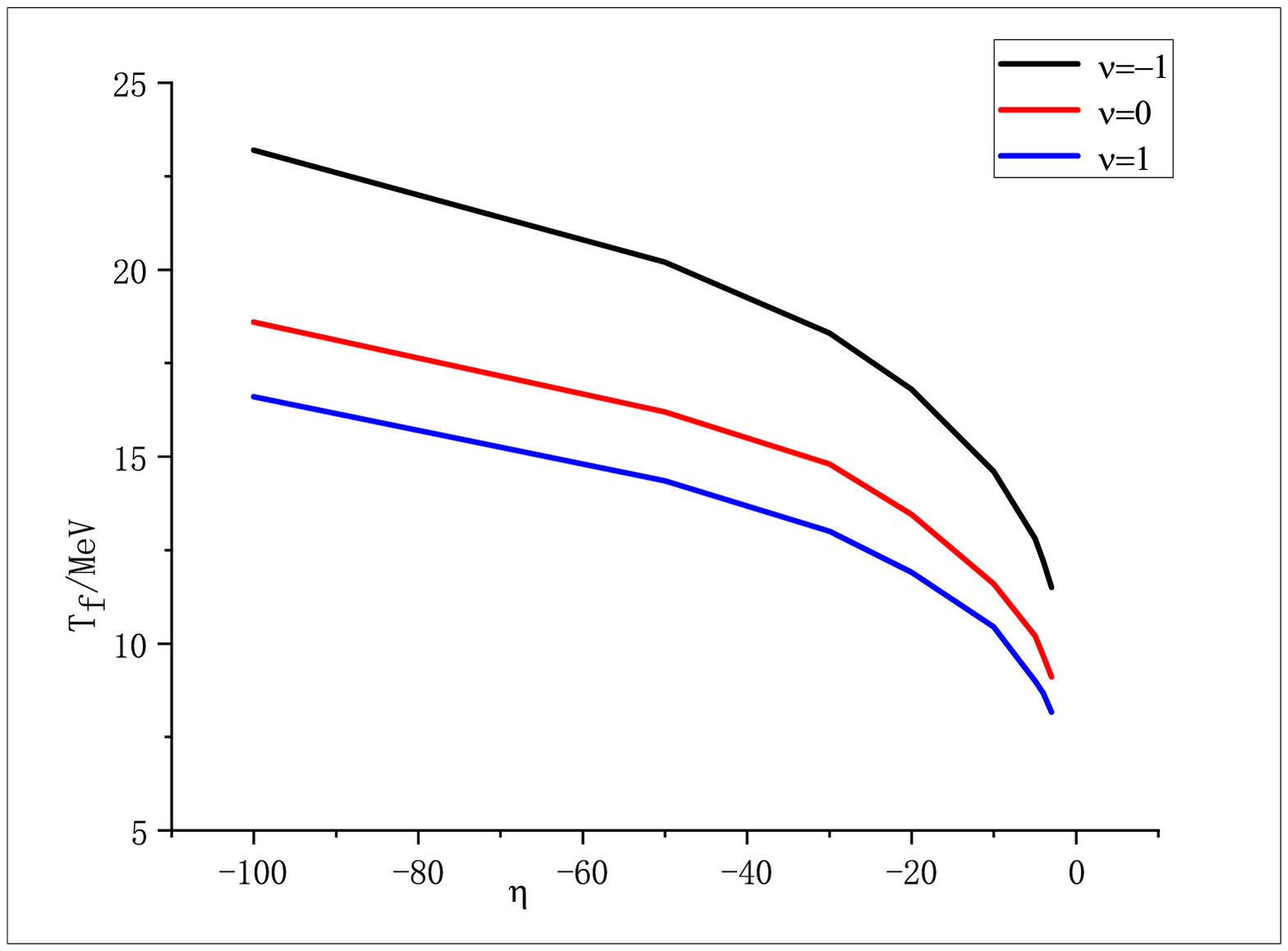}
 \includegraphics[scale=0.3]{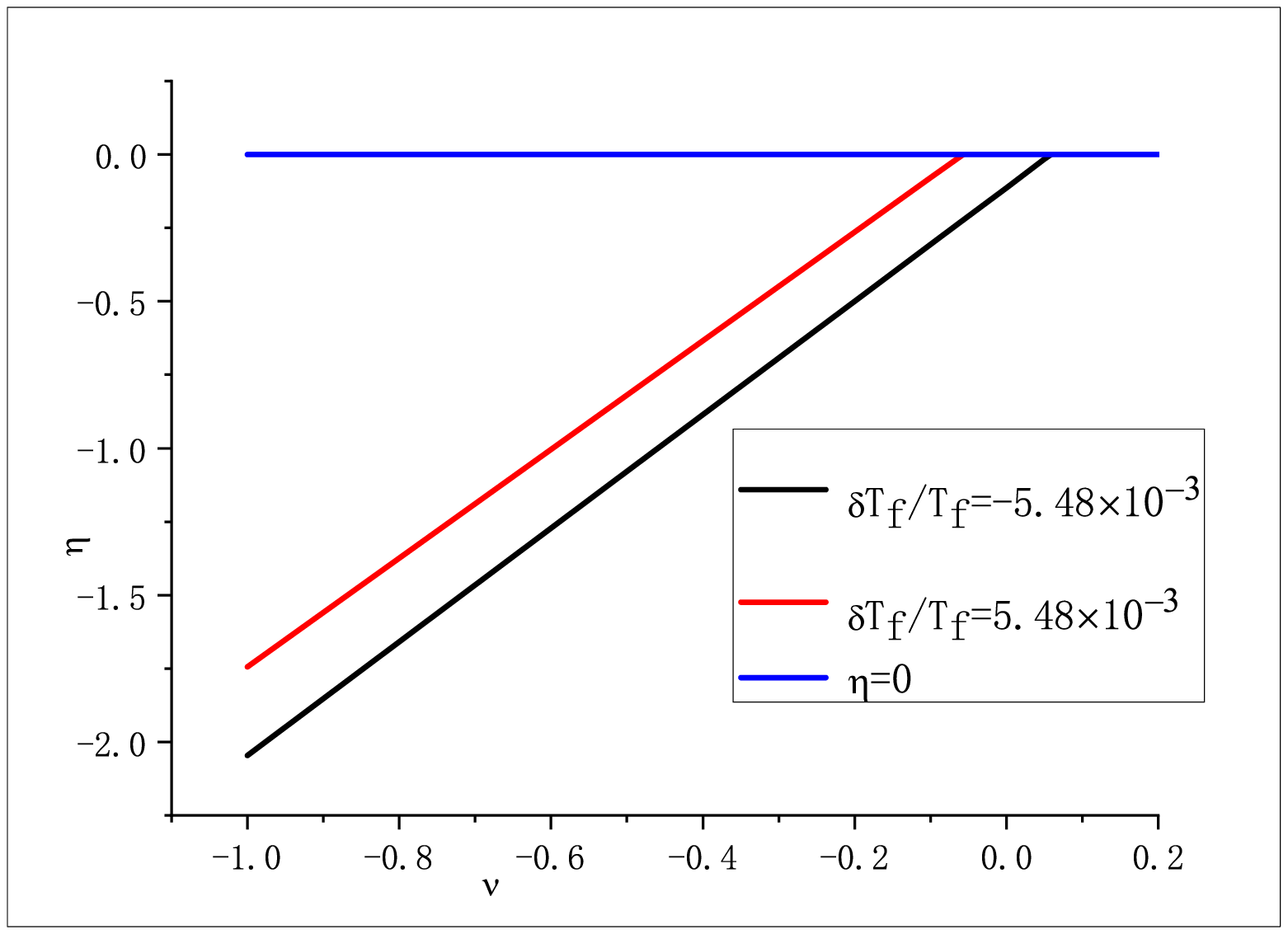}
\caption{The freeze-out temperature of the third Weyl type cosmological model as a function of the model parameter $\eta$ (left panel), and the freeze-out constraint (right panel), obtained from the numerical solution of the field equation. For the initial value of the Weyl vector we have adopted the value $\psi (0)=0$. }
\label{fig6}
\end{figure*}

\paragraph{The numerical freeze-out constraint.} The dependence of the freeze-out temperature on the parameters of the third Weyl type cosmological model and the freeze-out constraint on the model parameters are presented in Fig~\ref{fig6}.

\paragraph{The analytical freeze-out constrain.} By taking into account that for this model $f_Q=\left(\eta \nu/6\right)e^{\mu Q/\left(6H_0^2\right)}$, and $f_T=\nu/6\kappa ^2$, we obtain for the freeze-out constraint the general expression
\begin{eqnarray}
\left\vert \frac{\delta \mathcal{T}_{f}}{\mathcal{T}_{f}}\right\vert  &=&%
\frac{1}{10c_{q}}\sqrt{\frac{\pi ^{2}g_{\ast }}{180\kappa ^{2}}}\frac{1}{%
\mathcal{T}_{f}^{3}}\Bigg(1-\frac{30\kappa ^{2}}{\pi ^{2}g_{\ast }}\frac{%
\eta H_{0}^{2}e^{\frac{\mu Q}{6H_{0}^{2}}}+ \frac{\nu} {6\kappa ^{2}} T}{%
\mathcal{T}_{f}^{4}}\nonumber\\
&&+\frac{\eta \mu }{3}e^{\frac{\mu Q}{6H_{0}^{2}}} +\frac{4}{9}\nu +\frac{m^{2}}{12\kappa ^{2}}\Bigg).
\end{eqnarray}

Since $Q|_{\mathcal{T}_{f}}=6H^{2}|_{\mathcal{T}_{f}}\approx 6H_{GR}^{2}|_{%
\mathcal{T}_{f}}=\left( \pi ^{2}g_{\ast }/30\kappa ^{2}\right) \mathcal{T}%
_{f}^{4}$, by performing a first order expansion of the exponential, and by
neglecting the trace of the energy-momentum tensor, we obtain
\bea
\left\vert \frac{\delta \mathcal{T}_{f}}{\mathcal{T}_{f}}\right\vert &=&\frac{1%
}{10c_{q}}\sqrt{\frac{\pi ^{2}g_{\ast }}{180\kappa ^{2}}}\frac{1}{\mathcal{T}%
_{f}^{3}}\Bigg(1-\frac{30\kappa ^{2}\eta H_{0}^{2}}{\pi ^{2}g_{\ast }%
\mathcal{T}_{f}^{4}}-\frac{2}{3}\eta \mu \nonumber\\
&&+\frac{\pi ^{2}\eta \mu ^{2}g_{\ast
}}{30H_{0}^{2}\kappa ^{2}}\mathcal{T}_{f}^{4}+\frac{4}{9}\nu +\frac{m^{2}}{%
12\kappa ^{2}}\Bigg).
\eea

By neglecting the terms in the bracket containing the freeze-out temperature
$\mathcal{T}_{f}^{4}$, we obtain the following expression  on the model
parameters
\begin{equation}
1-\frac{2}{3}\eta \mu +\frac{4}{9}\nu +\frac{m^{2}}{12\kappa ^{2}}\approx
10c_{q}\sqrt{\frac{180\kappa ^{2}}{\pi ^{2}g_{\ast }}}\mathcal{T}%
_{f}^{3}\left\vert \frac{\delta \mathcal{T}_{f}}{\mathcal{T}_{f}}\right\vert
,
\end{equation}%
or
\begin{equation}
1-\frac{2}{3}\eta \mu +\frac{4}{9}\nu +\frac{m^{2}}{12\kappa ^{2}}<0.002.
\end{equation}

%\subsubsection{Deuterium and Helium constraints}

\subsubsection{The $^7 Li/H$ ratio}

\begin{figure*}[ht]
\centering
\includegraphics[scale=0.3]{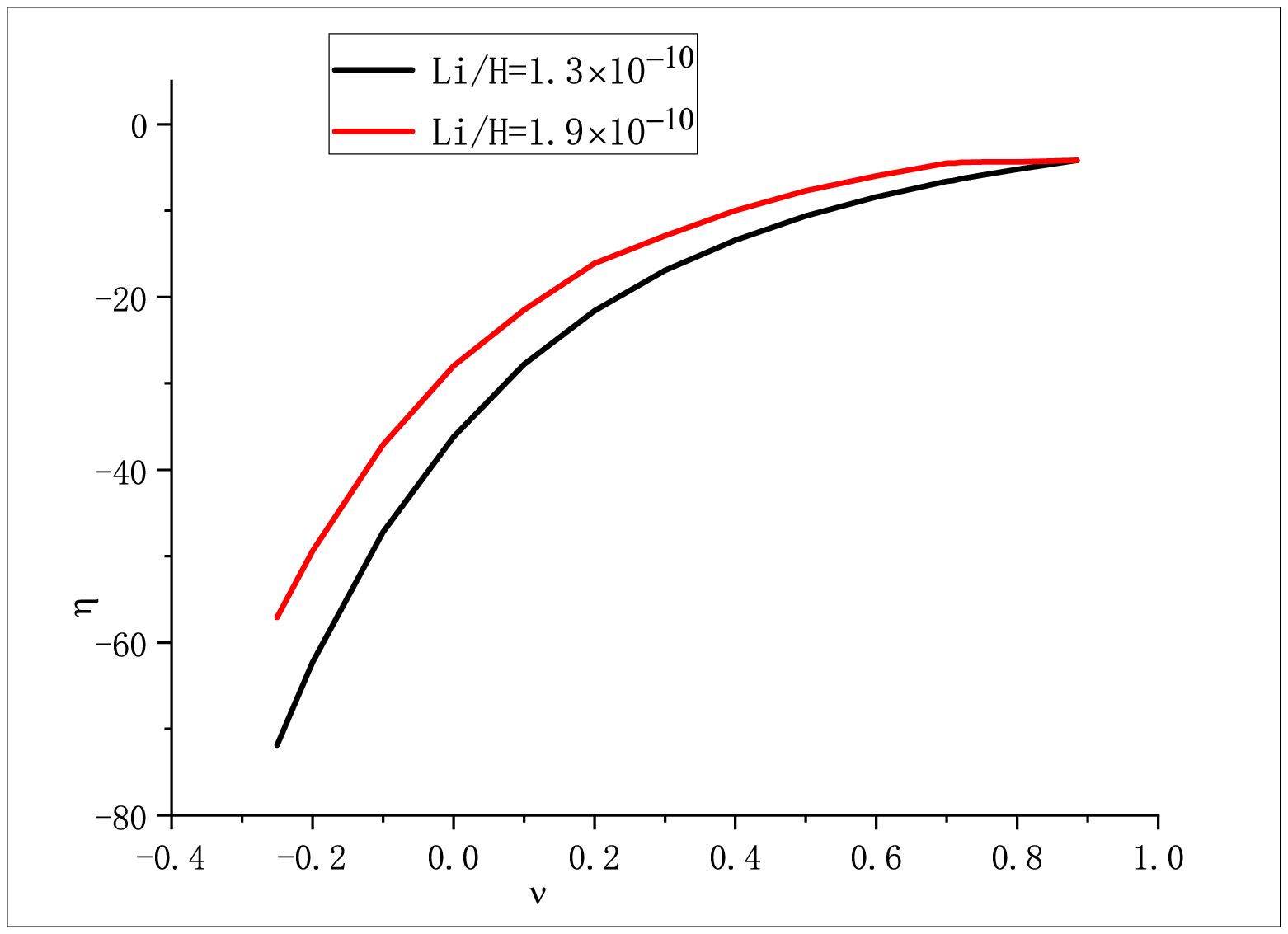}
\includegraphics[scale=0.3]{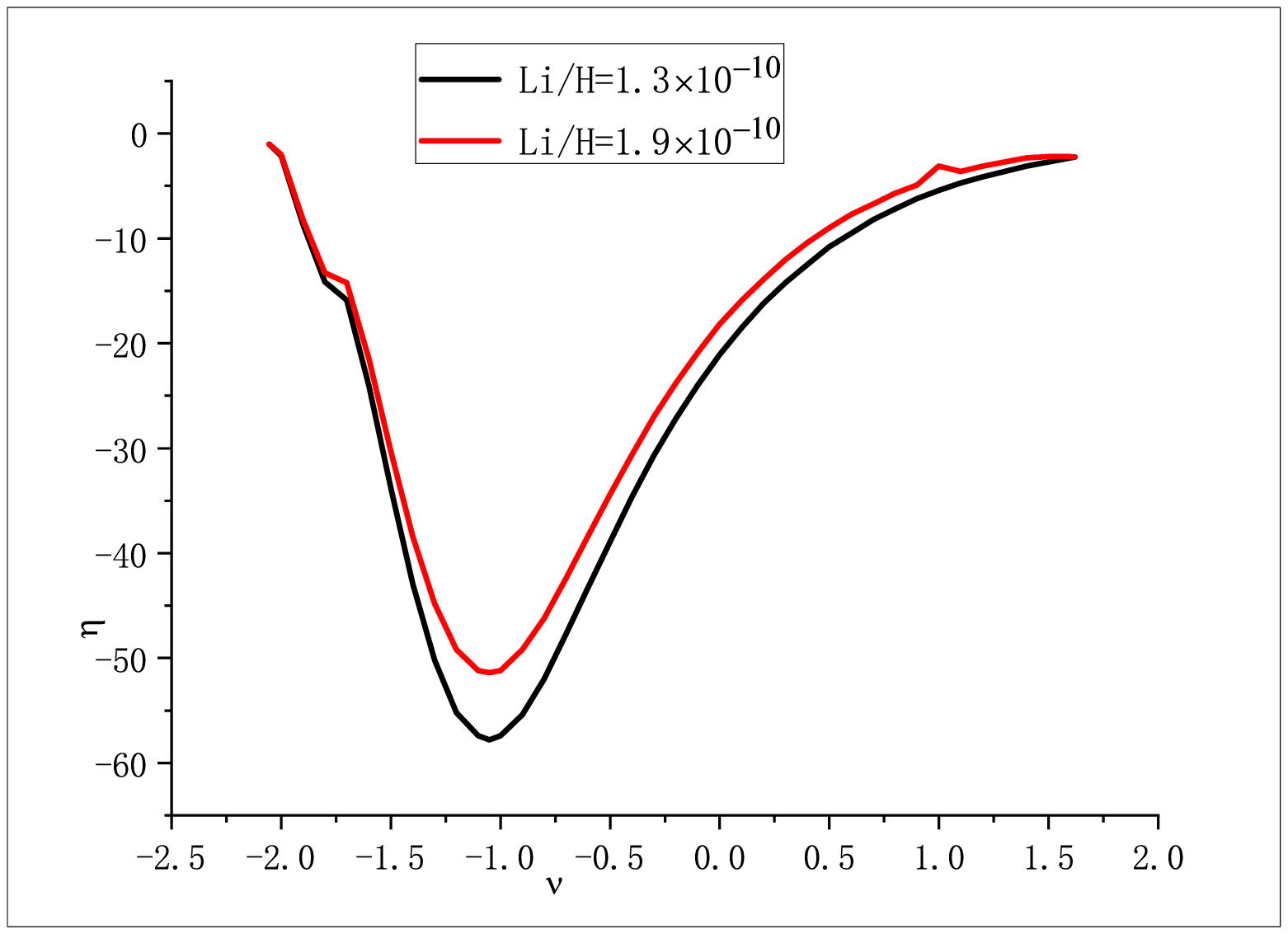}
\includegraphics[scale=0.3]{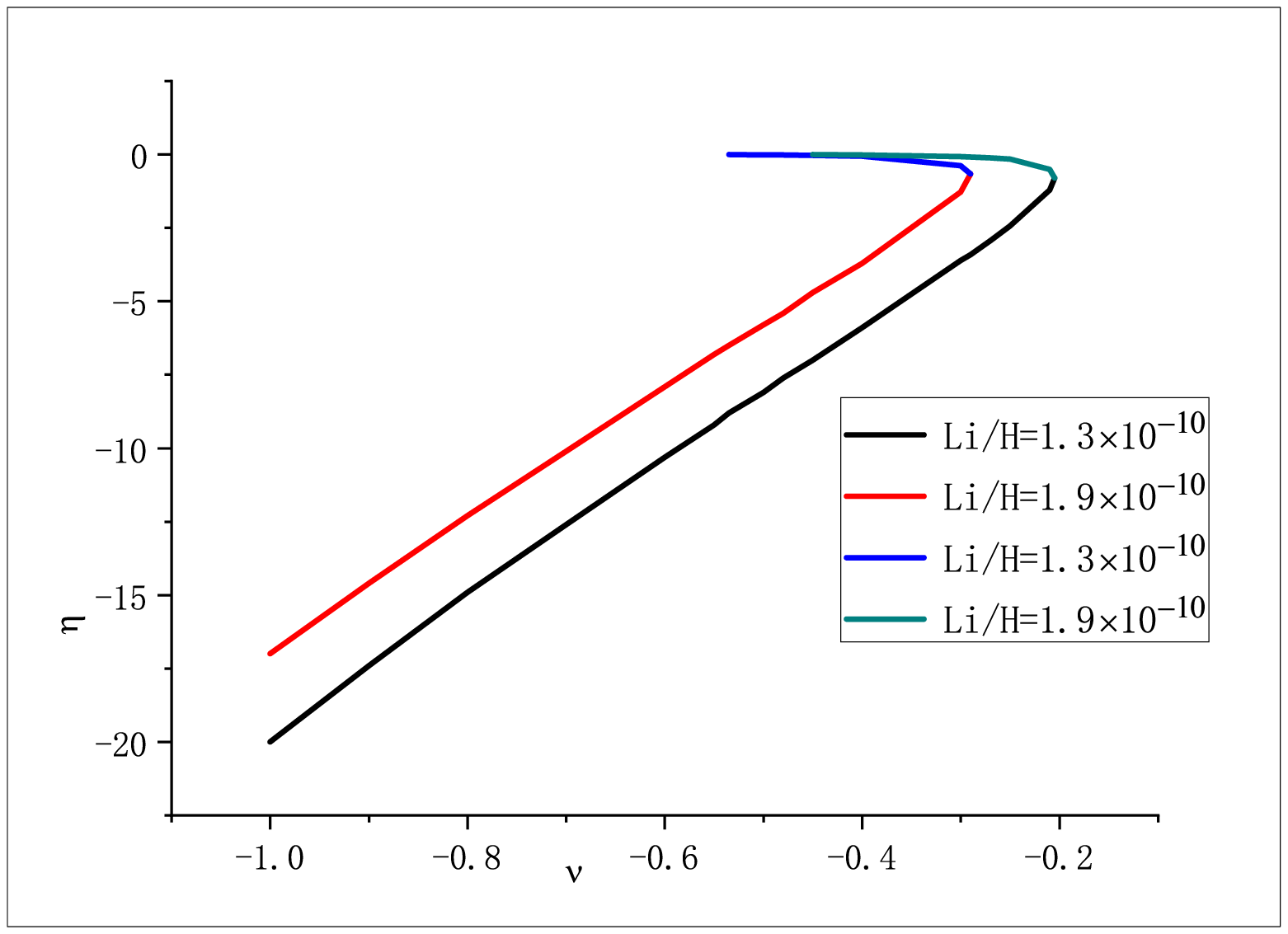}
 \includegraphics[scale=0.3]{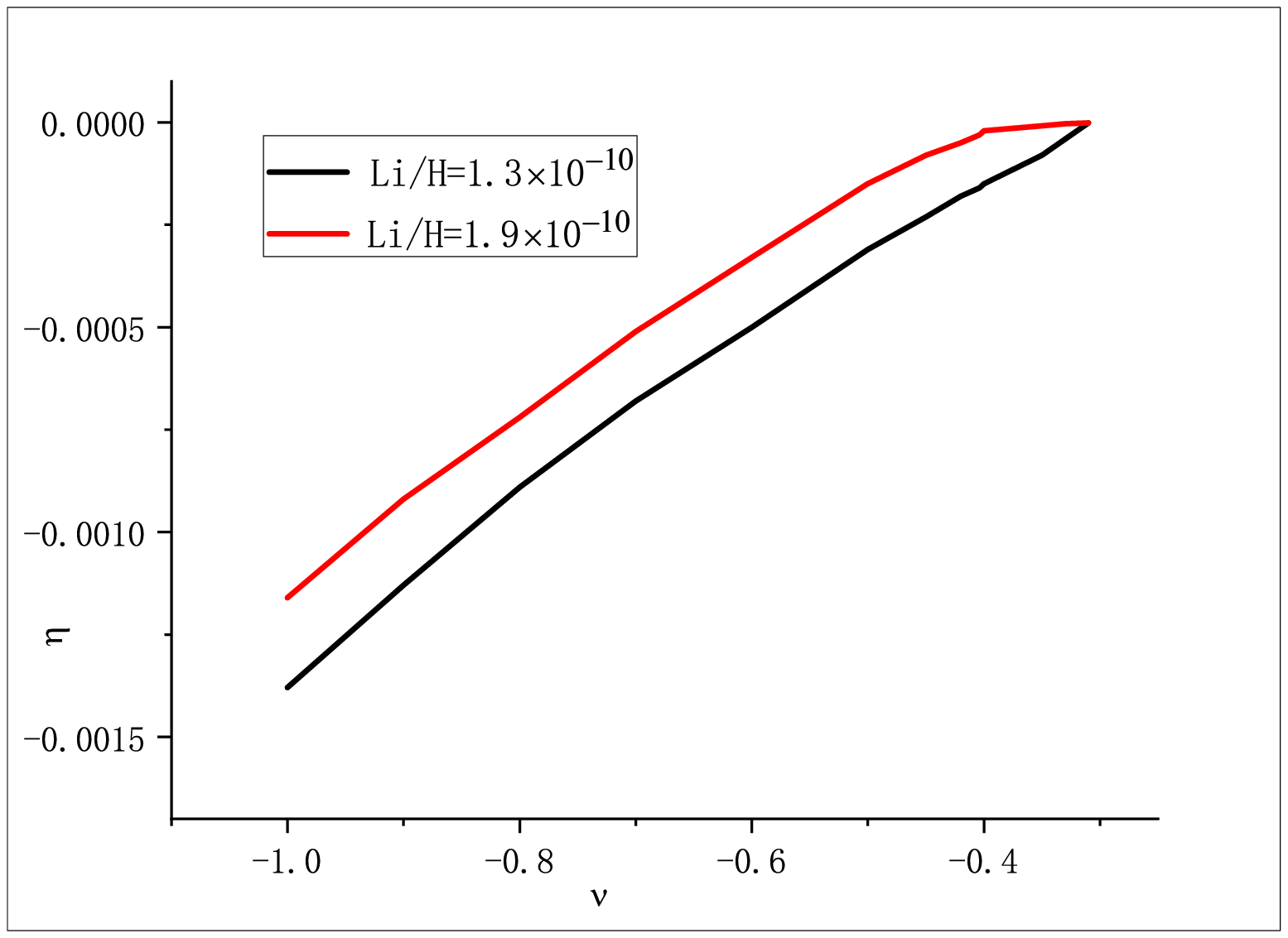}
\caption{The Lithium abundance constraint on the parameters $\eta$ and $\nu$ of the third Weyl type $f(Q,T)$ theory cosmological model. For the parameter $\mu$ w we have adopted the value $\mu =0.5$, while the initial conditions for the Weyl vector $\psi$ are $1\times 10^{-25}$ GeV,  $2\times 10^{-25}$ GeV,  $5\times 10^{-25}$ GeV, and $1\times 10^{-24}$ GeV, respectively.}
\label{fig7a}
\end{figure*}

The dependence of the $^7 Li/H$ ratio on the third Weyl type $f(Q,T)$ theory cosmological model on the parameters $\eta$ and $\nu$ is presented in Fig.~\ref{fig7a}.

\section{Discussions  and final remarks}\label{sect4}

General relativity is presently facing several observational challenges, mostly coming from the late time Universe observations.  There is a plethora of such observational data, obtained from the intensive investigation of the Cosmic Microwave Background, Supernovae Type Ia, Baryonic Acoustic Oscillations, Gamma-ray bursts, large scale structure growth data, Cosmic Chronometers etc. Even that some of these observations can be described by the standard $\Lambda$CDM model, several open problems confronting this model have become more and more important recently, like the Hubble or the $\sigma _8$ tensions \cite{Val}. This raises some new questions about the validity of standard cosmology, and of general relativity. And, of course, the problems of the cosmological constant and of the renormalizability of a possible quantum theory of gravity are still unsolved.

Hence, modified gravity theories can help in solving these problems, and they can open a new window on the gravitational dynamics. There are a large number of such theories, whose predictions sometimes significantly differ from those of standard general relativity. Among these theories those based on Weyl geometry are particularly interesting, since they could represent a bridge between gravity and elementary particle physics, due to their properties of conformal invariance.  The $f(Q)$ type theories are completely equivalent to GR for the choice $f(Q)=Q$, and for an arbitrary $f(Q)$ they can also describe the late-time properties of the cosmological expansion. An extension of the $f(Q)$ theory, the $f(Q,T)$ gravity, allows the existence of a coupling between matter and geometry, thus adding a new degree of freedom to the basic gravitational theory. A simpler formulation of the $f(Q,T)$ theory, which is closer to the spirit of the initial Weyl geometry, is represented by the Weyl type $f(Q,T)$ theory, in which the Weyl vector appears as an integral part of the gravitational dynamics. The applications of this theory in cosmology and astrophysics have pointed out its potential as an alternative to general relativity.

Modified gravity theories usually introduce a number of numerical parameters, which must be fine-tuned in order to explain the observational data. The most common method to constrain these parameters is by comparison with the late-time cosmological observations. However, many modified gravity theories that describe well the recent acceleration of the Universe do introduce some supplementary terms in the field equations that could significantly change the early time evolution. Therefore, it is also necessary to test modified gravity theories in the cosmological background of the early Universe,  no matter how successful they are in explaining the recent dynamics of the Universe.
late-universe successes that a modified gravity model may have, one should always examine whether the model can pass the BBN constraint

One of such phases in the evolution of the Universe, which can be successfully used as a testing ground for modified gravity theories, is the BBN phase. The concordance cosmology model is extremely successful in the description of the BBN phase, and thus it also offers the possibility of a detailed quantitative and qualitative comparison between general relativity and its alternatives. In the present work we have compared the predictions  of the Weyl type $f(Q,T)$ gravity with respect to the observational data of the BBN. We have adopted three specific functional forms of the function $f$, involving additive and multiplicative algebraic structures of the function $f$, as well as an exponential dependence on the nonmetricity scalar. We have investigated the range of the model parameters that allow to explain the abundances of the light elements in the early Universe, as well as the freeze-out temperature. For all our models we have also investigated numerically the early Universe evolution in the radiation dominated epoch, and we have obtained the temperature dependence of the Hubble function.

The first cosmological model we have considered has the Lagrangian density given by $f(Q,T)=\alpha Q+\left(\beta /6\kappa ^2\right)$, where $\alpha $ and $\beta$ are the model parameters. Recently, the cosmological implications of the Weyl type $f(Q,T)$ cosmological  models have been investigated in \cite{W19}. For the first model comparison with the observational data the values $\alpha =-1.07$ and $\beta =0.5$ have been adopted. These numerical values guarantee the positivity of the energy density, and a dark energy equation of state parameter that satisfies the observational constraints. These values also satisfy the freeze-out constrain, as one can see from Fig.~\ref{fig01}. The same range of values of $\alpha$ and $\beta$ also satisfies the Lithium constraint, as one can see from Fig.~\ref{fig02}. On the other hand, for the parameter $\beta$ taking values in the range $0.5<\beta<3$, the Strong Energy Condition (SEC) is violated by this model. However, for the first Weyl type $f(Q,T)$ gravity cosmological model, there is a range of the parameters $\alpha $ and $\beta$ satisfying both the SEC and the BBN constraints.

The late time behavior of  the second cosmological model we have analyzed, with $f(Q,T)$ given by $f(Q,T)=\left(\gamma /6H_0^2\kappa ^2\right)QT$, was also investigated in \cite{W19}, for $\gamma =0.315$. For the model the parameter of the dark energy equation of state $\omega$ lies in the range $-1<\omega <-1/3$, and hence the dark energy is of quintessence type. On the other hand, the Strong Energy Condition is violated at the present epoch for small variations of $\gamma$. Even for a small change in $\gamma$, the SEC doe nit hold anymore. On the other hand it is important to note that the NEC and the DEC are satisfied in this model. From the point of view of the BBN, all the observational constraints (freeze-out temperature, and light element abundances) can be satisfied by a range of the parameter $\gamma$, which is also consistent with the late time behavior observations.

In the present study, in order to constrain our theoretical models we have used the most stringent cosmological limits from BBN. In particular, we have used  we utilized the primordial Deuterium, Helium, and Lithium abundances to constrain the parameters of the three cosmological models in the Weyl type $f(Q,T)$ gravity.  The constraints obtained can be further improved once the new results of the future observations, like, for example, the CMB-S4 measurements \cite{S4} will become known.  At temperatures smaller than  0.1 MeV, an important reaction taking place in the early Universe is the production of deuterium via the reaction $n + p \leftrightarrow D +\gamma$.
The abundance of Deuterium decreases in the post-BBN Universe due to its annihilation during stellar evolution. Therefore BBN is the unique significant source of this element.  By using the primordial abundances of deuterium, we have obtained a set of constraints on the free parameters of the Weyl type $f(Q,T)$ gravity cosmological models. Deuterium is the most constrained light nucleus  among all light elements, with its observational determination reaching 1\% accuracy. Recently, the EMPRESS collaboration has presented  a measurement of the primordial $^4$He abundance that is over 3$\sigma$ from the $\Lambda$CDM theoretical value. A small 2$\sigma$ discrepancy may also exist in the deuterium abundance. These observational results strongly suggest that the $\Lambda$CDM paradigm may face some problems not only in the late Universe, but also during the BBN phase. Hence, modified gravity theories could play an important role in elucidating not only the cosmological behavior of the late Universe, but also in the description of the very early phases of cosmological evolution. In the present study we have considered the effects of such a modified gravity theory on the BBN phase, which could open a new perspective on the physics and cosmology of the early Universe.

%Although there is lithium problem in the standard BBN scenario, it can be different in other models.We have attempted to solve it using the Weyl $f(Q,T)$ gravity theory.As %we can see,in the first and the second model,Li problem can be compromised with the He abundance.

\section*{Acknowledgments}

 HHZhang is supported by the National Natural Science Foundation of China under Grant No. 12275367.
LMing acknowledges the Project funded by China Postdoctoral Science Foundation under Grant No. 2022M723677. This work is also supported by  the Fundamental Research Funds for the Central Universities, the Natural Science Foundation of Guangdong Province, and the Sun Yat-Sen University Science Foundation.

\end{document}